\documentclass[aps,jcp,twocolumn,groupedaddress]{revtex4}
\usepackage{graphicx}
\usepackage{amsmath}
\usepackage{dcolumn}
\usepackage{bm}
\usepackage{ulem}
\usepackage{amssymb}
\usepackage{epsfig}
\usepackage{xcolor}

\newcommand{\rmd}{\mathrm{d}}
\newcommand{\rme}{\mathrm{e}}
\newcommand{\rmRe}{\mathrm{Re}}

\newcommand{\ocal}{\mathcal{O}}
\newcommand{\ccal}{\mathcal{C}}
\newcommand{\zcal}{\mathcal{Z}}
\newcommand{\lcal}{\mathcal{L}}
\newcommand{\bcal}{\mathcal{B}}
\newcommand{\phih}{\hat{\phi}}
\newcommand{\cphi}{\varphi}
\newcommand{\muh}{\hat{\mu}}
\newcommand{\taua}{{\tau_\alpha}}

\begin{document}
\title{Generalized mode-coupling theory of the glass transition: schematic results at finite and 
infinite order}

\author{Liesbeth M.~C.~Janssen}
\altaffiliation{Current address: Institute for Theoretical Physics II,
                Heinrich-Heine University D\"usseldorf, Universit\"atsstra{\ss}e 1, D-40225, Germany}
\email[Electronic mail: ]{ljanssen@thphy.uni-duesseldorf.de}
\affiliation{Department of Chemistry, Columbia University, 3000 Broadway, New York, New York 10027, USA}
\author{Peter Mayer}
\altaffiliation{Current address: ThinkEco, Inc., 494 8th Avenue, New York, NY 10001, USA}
\email[Electronic mail: ]{peter@thinkecoinc.com}
\affiliation{Department of Chemistry, Columbia University, 3000 Broadway, New York, New York 10027, USA}
\author{David R.~Reichman}
\email[Electronic mail: ]{drr2103@columbia.edu}
\affiliation{Department of Chemistry, Columbia University, 3000 Broadway, New York, New York 10027, USA}

\date{\today}

\begin{abstract}
We present an extensive treatment of the generalized mode-coupling theory
(GMCT) of the glass transition, which seeks to describe the dynamics of
glass-forming liquids using only static structural information as input.  This
theory amounts to an infinite hierarchy of coupled equations for multi-point
density correlations, the lowest-order closure of which is equivalent to
standard mode-coupling theory.  Here we focus on simplified schematic GMCT
hierarchies, which lack any explicit wavevector-dependence and therefore allow
for greater analytical and numerical tractability.  For one particular
schematic model, we derive the unique analytic solution of the infinite
hierarchy, and demonstrate that closing the hierarchy at finite order leads to
uniform convergence as the closure level increases. We also show numerically
that a similarly robust convergence pattern emerges for more generic schematic
GMCT models, suggesting that the GMCT framework is generally convergent, even
though no small parameter exists in the theory.  Finally, we discuss how
different effective weights on the high-order contributions ultimately control
whether the transition is continuous, discontinuous, or strictly avoided,
providing new means to relate structure to dynamics in glass-forming systems.
\end{abstract}

\maketitle

\section{Introduction}
Understanding the behavior of supercooled liquids and the process of glass
formation represents one of the major challenges in condensed matter physics
\cite{anderson:95,debenedetti:01,berthier:11}. Glass-forming systems exhibit a
dramatic slowdown of the dynamics upon mild supercooling or compression, but at
the same time undergo only subtle structural changes at the atomic level. The
question how such marginal differences in structure can be accompanied by an
orders-of-magnitude change in the relaxation dynamics lies at the heart of the
glass-transition problem.  While several plausible theoretical frameworks have
been developed to rationalize this behavior
\cite{cavagna:09,berthier:11,tarjus:11, biroli:13,langer:14}, there is still no
fully microscopic theory available that can accurately describe the dynamics of
supercooled liquids over all relevant time and temperature scales.

Among the various frameworks proposed in the last few decades, the mode-coupling
theory (MCT) of the glass transition stands out as the only theory of glassy
dynamics based entirely on first principles \cite{gotze:92,gotze:09}.  MCT seeks to predict the
dynamics of the time-dependent density-density correlation function $F(k,t)$ at
wavelength $k$ and time $t$ using only time-independent structural
properties, such as the static structure factor $S(k) \equiv F(k,0)$, as input.
Within the MCT framework, the dynamics of $F(k,t)$ is governed by a memory
function involving (projected) dynamic \textit{four}-point correlation functions, which
probe correlations between pair-density modes. The crucial MCT approximation is
to factorize these four-point correlations into products of $F(k,t)$'s, thus
rendering a closed, self-consistent theory. The most notable success of MCT is that
it correctly predicts the growth of a plateauing region in $F(k,t)$ upon supercooling,
and it offers a simple physical picture for the initial phase of the dynamic slow-down
in terms of the cage effect \cite{reichman:05,gotze:09}.
With rescaling of temperature or density, the theory can also be made 
quantitatively successful in the weakly to moderately supercooled regime \cite{weysser:10}.
However, the uncontrolled nature of the factorization approximation
ultimately breaks down, leading to the prediction of a
spurious dynamical transition at temperatures well above the experimentally
observed glass transition (see for reviews e.g.\ \cite{kob:03,reichman:05,szamel:13}).

A promising approach to improve upon standard MCT was first introduced by
Szamel in 2003 \cite{szamel:03}. This approach, which we refer to as
generalized mode-coupling theory (GMCT), relies on the fact that the exact time
evolution of the four-point correlations is governed by six-point correlations,
which in turn are controlled by eight-point correlations, and so on.  This
leads to a \textit{hierarchy} of coupled equations that makes it possible to
delay the uncontrolled factorization approximation to a later stage. Szamel
\cite{szamel:03} and Wu and Cao \cite{wu:05} showed that applying the
factorization at the level of six- or eight-point correlations indeed
systematically improves the predicted location of the dynamical transition.
In a subsequent proof-of-principle study, two of us \cite{mayer:06} extended
this approach to \textit{infinite} order using a simplified schematic
(wavevector-independent) model.  Remarkably, it was found that this infinite
hierarchy of schematic GMCT equations admits an analytical solution that rounds
off the MCT transition completely, and converts the power-law divergences of 
transport coefficients into activated-like behavior.  More recently, we found that infinite-order
schematic GMCT can in fact account for both strictly avoided and sharp standard-MCT-like 
transitions, depending on the choice of the level-dependent schematic
parameters \cite{janssen:14}.  We also identified a class of schematic
hierarchies out of which the concept of fragility, i.e.\ the degree to which
the relaxation-time growth deviates from Arrhenius behavior \cite{angell:95},
emerges naturally.  This is to be contrasted with standard MCT, which can only
predict ''fragile" relaxation and hence fails to distinguish between materials
with different fragilities.  Finally, it was recently shown that the fully
microscopic version of the theory, which contains no free parameters and takes
only static structural information as input, already gives unprecedented
accuracy for the time-dependent dynamics of a realistic quasi-hard sphere
system when only the lowest few levels of the hierarchy are considered
\cite{janssen:15}. More specifically, full quantitative agreement could be obtained 
between GMCT and computer simulations in the mildly supercooled regime,
without having to resort to density or wavevector rescaling.
 These studies all suggest that GMCT offers a promising and
highly versatile framework to describe the behavior of glass-forming matter.
Since GMCT can \textit{systematically} correct upon standard MCT by 
incorporating increasingly many higher-order correlations, the theory thus 
poses an appealing and potentially pragmatic route towards an ultimately fully
quantitative and fully microscopic description of glassy dynamics over all relevant time and
temperature domains.

Despite these encouraging results, it is not \textit{a priori} clear that the
GMCT approach, which seeks to systematically defer the uncontrolled MCT
factorization, is in itself controlled. Indeed, a justified concern is whether
the GMCT hierarchy converges at all, since there exists no small parameter in
the theory that would warrant higher-order contributions less important. In
this work, we seek to address this question using both analytical and numerical
arguments. For simplicity, we will restrict our discussion to the case of
schematic GMCT hierarchies, but we expect our qualitative results and
conclusions to apply to the microscopic theory as well. The model of Ref.\
\cite{mayer:06}, which we will refer to here as the Mayer-Miyazaki-Reichman
model, offers an ideal test ground for this purpose, since a unique analytic
solution is at hand for the infinite-order limit. For completeness, we shall
first present the full derivation of this highly non-trivial solution (the
final result of which was presented in Ref.\ \cite{mayer:06}), and subsequently
consider various \textit{finite}-level closures of this hierarchy. It will be
shown rigorously, through a rather lengthy derivation, that the finite-order
solutions converge systematically towards the analytic infinite-order result as
the closure level tends to infinity. We then provide numerical evidence that
a similar convergence pattern also applies to more generic GMCT hierarchies,
such that the inclusion of more hierarchical levels directly translates
into accurate dynamics over increasingly long time scales.
This suggests that the GMCT approach is indeed controlled in the sense that,
\textit{independent of closure method}, the dynamics described by a finite level
of the hierarchy coincides with that of the complete hierarchy for an
ever increasing time duration. 
 Finally, we seek to
provide more physical insight into how the influence of higher-order
correlations dissipates through the GMCT hierarchy to ultimately control the
dynamics of $F(k,t)$. To this end, we will make a connection between infinite
schematic hierarchies and the standard-MCT $F_2$ model
\cite{leutheusser:84,bengtzelius:84}, and discuss how the functional forms of
the level-dependent schematic parameters control whether the transition is
continuous (''type-A"), discontinuous (''type-B"), or strictly avoided.
Overall, this work can help to assess the success of GMCT as a
first-principles-based theory of the glass transition.

The layout of this paper is as follows.  We first review the fully microscopic
GMCT equations in Sec.\ \ref{sec:GMCT} and subsequently describe how the theory
may be reduced to more tractable schematic models.  In Sec.\
\ref{sec:MMRmodel}, we then give an extensive treatment of the
Mayer-Miyazaki-Reichman model of Ref. \cite{mayer:06}, which constitutes one of
the simplest possible schematic GMCT hierarchies. We will first present the
general solution to this model in Sec.\ \ref{sec:gensol}, and then rigorously
derive its analytic solution in the infinite-order limit (Sec.\ \ref{sec:inf}).
Next, we provide a detailed discussion of so-called mean-field closures to
close the Mayer-Miyazaki-Reichman hierarchy at finite order (Sec.\
\ref{sec:mfn}), and subsequently consider alternative ways to truncate the
hierarchy (Sec.\ \ref{sec:altc}).  In Sec.\ \ref{sec:genericsGMCT}, we then
consider more generic functional forms of schematic GMCT hierarchies, which
allow for more flexibility in the choice of schematic parameters and which can
account for a much broader pallet of glassy relaxation behaviors.  We first
discuss their general convergence patterns in Sec.\ \ref{sec:genconv}, and
finally explore how the higher-order correlations may ultimately control the
physical nature of the predicted glass transition in Sec.\
\ref{sec:transtypes}.  Concluding remarks are given in Sec.\ \ref{sec:concl}.

\section{Generalized mode-coupling theory equations} 
\label{sec:GMCT}

In this section, we provide the physical and mathematical background for the
infinite hierarchy of GMCT equations. We first summarize the results of the
full wavevector-dependent, time-dependent theory \cite{janssen:15}, which
contains no free parameters and is based entirely on first principles. Next, we
discuss how these fully microscopic equations may be reduced to simpler
schematic models that allow for more analytical and numerical tractability,
and that form the basis for the remainder of this paper.

\subsection{Microscopic (\bm{$k$}-dependent) equations of motion}

The microscopic GMCT equations with full time- and wavevector-dependence 
were first derived in Ref.\ \cite{janssen:15},
and here we briefly recall the main results of this theory. We consider the
dynamics of the normalized $2n$-point
density correlation functions $\Phi_n(k_1,\hdots,k_n, t)$, which probe particle
correlations over $n$ distinct $k$-values,
\begin{equation}
\label{eq:Phindef}
\Phi_n(k_1,\hdots,k_n, t) = \frac{\langle \rho_{\mathbf{-k_1}}(0) \hdots \rho_{-\mathbf{k_n}}(0)
                                  \rho_{\mathbf{k_1}}(t) \hdots \rho_{\mathbf{k_n}}(t) \rangle}
                                 {\langle \rho_{\mathbf{-k_1}}(0) \hdots \rho_{-\mathbf{k_n}}(0)  
                                  \rho_{\mathbf{k_1}}(0) \hdots \rho_{\mathbf{k_n}}(0) \rangle}.
\end{equation} 
Within the GMCT framework, these correlators obey the general equations of
motion
\begin{gather} 
\ddot{\Phi}_n(k_1,\hdots,k_n,t) + \zeta\dot{\Phi}_n(k_1,\hdots,k_n,t) 
\nonumber \\ 
+ \Omega^2_n(k_1,\hdots,k_n) \Phi_n(k_1,\hdots,k_n,t) 
\nonumber \\ 
+\int_0^t M_n(k_1,\hdots,k_n,\tau) \dot{\Phi}_n(k_1,\hdots,k_n,t-\tau) d\tau  = 0. \label{eq:GMCTeqPhi_n} 
\end{gather} 
The label $n$ ($n=1,\hdots,\infty$) thus specifies the level of the hierarchy.
In Eq.\ (\ref{eq:GMCTeqPhi_n}), the dots denote time derivatives, $\zeta$ represents an
effective friction coefficient that accounts for the short-time dynamics, and the bare frequencies
are given by 
\begin{equation} 
\label{eq:Omega2n}
\Omega^2_n(k_1,\hdots,k_n) = \frac{k_{\rm{B}}T}{m} \left[ \frac{k_1^2}{S(k_1)} + \hdots + \frac{k_n^2}{S(k_n)} \right]. 
\end{equation}
Here $k_{\rm{B}}$ is the Boltzmann constant, $T$ is the temperature, and $m$ is
the particle mass.
 For the memory functions we have
\begin{eqnarray}
\label{eq:Mn} 
M_n(k_1,\hdots,k_n,t) = \frac{\rho k_{\rm{B}}T}{16m\pi^3} \sum_{i=1}^n \frac{\Omega^2_1(k_i)}{\Omega^2_n(k_1,\hdots,k_n)}
                  \nonumber \\
                   \times \int d\mathbf{q} |\tilde{V}_{\mathbf{q,k}_i-\mathbf{q}}|^2 
                   S(q) S(|\mathbf{k}_i-\mathbf{q}|)
                   \hphantom{XXXX}
                  \nonumber \\
                  \times \Phi_{n+1}(q,|\mathbf{k}_1-\mathbf{q}\delta_{i,1}|,\hdots,|\mathbf{k}_n-\mathbf{q}\delta_{i,n}|,t), \nonumber \\
\end{eqnarray} 
where $\rho$ is the total density, $\delta_{i,j}$ is the Kronecker delta
function, and $\tilde{V}_{\mathbf{q,k}_i-\mathbf{q}}$ are static vertices that represent
wavevector-dependent coupling strengths for the higher-order correlations.
These vertices are defined as
\begin{equation}
\tilde{V}_{\mathbf{q,k-q}} = 
 ({\hat{\mathbf{k}}} \cdot \mathbf{q}) c(q) + 
  {\hat{\mathbf{k}}} \cdot (\mathbf{k-q}) c(|\mathbf{k-q}|),
\end{equation}
where $\hat{\mathbf{k}} = \mathbf{k}/k$ and $c(q)$ denotes the direct correlation
function, which is related to the static structure factor as $c(q) \equiv
[1-1/S(q)] / \rho$ \cite{hansen:06}. The latter serves as the \textit{only 
input} to the theory (aside from trivial system parameters such as $\rho$ and $T$). 
Clearly, the hierarchical nature of the GMCT equations arises from the
memory kernel: for any given level $n$, the memory function contains $n$ different
dynamic 2($n+1$)-point correlators.
It should be noted that the time dependence of these dynamic 2($n+1$)-point
correlators occurs in a subspace \textit{orthogonal} to the $n$-point
density-mode basis that spans the 2$n$-point correlation functions.  However,
one may always project the $n$-point density components out of the multilinear
($n+1$)-point density basis, so that the $2(n+1)$-point
correlation functions evolve in the orthogonal subspace \textit{by construction} \cite{schofield:92}.
Hence, it is sensible to treat the evolution with ``normal" dynamics.
The remaining approximations in our framework
are the neglect of 
the off-diagonal 
dynamic $2n$-point correlation functions $\langle \rho_{-\mathbf{k}_1}(0) \hdots \rho_{-\mathbf{k}_n}(0)
\rho_{\mathbf{k}'_1}(t) \hdots \rho_{\mathbf{k}'_n}(t) \rangle$ with $\mathbf{k}_i \neq \mathbf{k}'_i$,
and the use of Gaussian and convolution approximations for the statics.
Note that these approximations are also commonly invoked at the standard-MCT level.
In principle, one could relax both of these approximations, i.e., incorporate more off-diagonal
dynamic correlators and keep all static correlators in their explicit multi-point form. 
Such higher-order statics would then serve as additional input to the theory.
The GMCT equations of motion of Eq.\ (\ref{eq:GMCTeqPhi_n}) are subject to the
boundary conditions $\Phi_n(k_1,\hdots,k_n,0)=1$ and $\dot{\Phi}_n(k_1,\hdots,k_n,0)=0$
for all $n$.


\subsection{Schematic equations of motion}
\label{seC:sGMCT}

Our schematic GMCT equations involve several simplifications of the microscopic
hierarchy presented in the previous section.  
Following Refs.\ \cite{mayer:06, janssen:14}, we drop the wavevector indices and treat
all wavevectors on an equal footing.  That is, $\Phi_1(k,t) \mapsto
\phi_1(t)$, $\Phi_2(k_1,k_2,t) \mapsto \phi_2(t)$, \dots, and
$\Omega^2_n(k_1,\hdots,k_n) \mapsto \mu_n$.  We also replace $\frac{\rho
k_{\rm{B}}T}{16m\pi^3} \sum_{i=1}^n \int d\mathbf{q}
|\tilde{V}_{\mathbf{q,k}_i-\mathbf{q}}|^2 S(q) S(|\mathbf{k}_i-\mathbf{q}|)$ by a level-dependent
constant $n \lambda_n$, which represents the effective weight of the
memory kernel at level $n$. This brings the memory functions into the form
$M_n(t) \mapsto \lambda_n \phi_{n+1}(t)$. Finally, we assume that the density
correlation functions decay so slowly that the second time-derivatives can be
neglected (i.e., overdamped dynamics),
and we set the friction constant $\zeta$ 
to 1. Note that the latter simply amounts to a rescaling of time.
Under these assumptions, we arrive at the generic schematic hierarchy  
\begin{equation}
\label{eq:sGMCThierarchy}
\dot{\phi}_n(t) + \mu_n \phi_n(t) + \lambda_n \int_0^t \phi_{n+1}(\tau) \dot{\phi}_n(t-\tau) d\tau = 0.
\end{equation}
This concludes our description of the schematic GMCT
equations which, although microscopically motivated, lack any explicit
$\mathbf{k}$-dependence. It is evident that Eq.\ (\ref{eq:sGMCThierarchy})
represents an infinite hierarchy of coupled equations, i.e., the time evolution
of any $\phi_n(t)$ is governed by $\phi_{n+1}(t)$, which in turn is governed by
$\phi_{n+2}(t)$, etc. Equation (\ref{eq:sGMCThierarchy}) is subject to the initial conditions
$\phi_n(0) = 1$ for all $n$. 
In the remainder of this paper, we shall consider several different choices 
for the $\mu_n$ and $\lambda_n$ parameters, and discuss the solutions
of the corresponding hierarchies at both finite and infinite order.

\section{The Mayer-Miyazaki-Reichman model: 
$\mu_n=n$ and $\lambda_n=\rm{constant}$}
\label{sec:MMRmodel}

In the work of Ref.\ \cite{mayer:06}, which constitutes the first study of schematic
GMCT, the authors considered the infinite hierarchy (\ref{eq:sGMCThierarchy})
with $\mu_n=\mu n$ and $\lambda_n = \Lambda$ (i.e., a constant).
The linear form of $\mu_n$ follows naturally from the microscopic
frequencies [Eq.\ (\ref{eq:Omega2n})], provided that no explicit distinction is made between
different wave vectors. Setting $\mu=1$ then yields: 
\begin{equation}
\label{equ:MCTdef}
  \dot{\phi}_n(t) + n \phi_n(t) + \Lambda \int_0^t \rmd\tau \, 
  \phi_{n+1}(t-\tau) \, \dot{\phi}_n(\tau) = 0. 
\end{equation}
We will refer to this hierarchy as the Mayer-Miyazaki-Reichman model.
Note that $\mu=1$ merely involves a rescaling of time, while the condition
$\lambda_n = \Lambda$ implies a \textit{level-independent} coupling of the
higher-order density modes. The control parameter $\Lambda$ may be physically interpreted
as e.g.\ an inverse-temperature-like or density-like parameter.

In this section, we will first explicitly derive the analytic
solution of the Mayer-Miyazaki-Reichman model in the infinite-order limit 
-- the final result of which is given in Eq.\ (\ref{equ:phihinf}) --,
and subsequently discuss the solutions under various finite-order closures. To this end,
we first introduce a generic closure function $\mathcal{C}$ that terminates the hierarchy at level
$N \geq 2$,
\begin{equation}
  \phi_N(t) = \ccal(\{\phi_n(t)\},t).
  \label{equ:cdef}
\end{equation}
Equations (\ref{equ:MCTdef}) and (\ref{equ:cdef}) thus constitute a closed
system of integro-differential equations governing the time evolution of the
functions $\{\phi_n(t)\} = \{\phi_1(t), \phi_2(t), \dots, \phi_N(t) \}$.  Note
that for $N=2$ and the closure $\phi_2(t) = \phi_1^2(t)$, we essentially
recover the standard-MCT schematic model of Leutheusser \cite{leutheusser:84}.

\subsection{General solution}
\label{sec:gensol}

Equation (\ref{equ:MCTdef}) relates any function $\phi_n(t)$ in the hierarchy
to $\phi_N(t)$. In this subsection we analyze this relation and derive its explicit form.
We first Laplace transform Eq.\ (\ref{equ:MCTdef}) as it is thereby reduced to an
algebraic equation. 
We use the standard Laplace
transform $\hat{f}(s) = \lcal\{ f(t) \}$ defined by
\begin{eqnarray}
  \lcal\{f(t)\} &=& \int_0^\infty \rmd t \, f(t) \, \rme^{-s t}, \nonumber \\ 
  \lcal^{-1}\{ \hat{f}(s) \} &=& \int_{c - i \infty}^{c + i \infty} \frac{\rmd s}{2 \pi i} \, \hat{f}(s) 
  \, \rme^{s t}. 
  \label{equ:Ldef}
\end{eqnarray}
%
%
For the inversion integral, $c$ must be chosen such that the integration contour is contained in the
half-plane where $\hat{f}(s)  = \lcal\{ f(t) \}$ converges and is analytic. Using standard properties
of Laplace transforms and the initial conditions $\phi_n(0) = 1$, one shows that Eq.\ (\ref{equ:MCTdef})
turns into
\begin{equation}
  \phih_n(s) = \frac{1}{\displaystyle{s+\frac{n}{1+\Lambda \phih_{n+1}(s)}}}. 
  \label{equ:contfrac}
\end{equation}
By iteration of Eq.\ (\ref{equ:contfrac}) we can, in principle, express any function $\phih_n(s)$
in terms of $\phih_N(s)$. However, this generates a continued-fraction type expression which
is not suitable as such for further analysis. A transformation that corresponds
to resolving nested fractions is
\begin{equation}
  \phih_n(s) = \frac{1}{s} \left[ 1 - n \, \frac{\cphi_{n+1}(s)}{\cphi_n(s)} \right], 
  \label{equ:mapfrac}
\end{equation}
which defines the new set of functions $\{ \cphi_n(s) \} = \{\cphi_1(s), \cphi_2(s), \ldots, 
\cphi_{N+1}(s) \}$. In fact, since Eq.\ (\ref{equ:mapfrac}) depends on ratios of the $\cphi_n(s)$
these are only defined up to an overall factor. Via substitution of Eq.\ (\ref{equ:mapfrac}) into
Eq.\ (\ref{equ:contfrac}) one verifies that the $\cphi_n(s)$ satisfy the \textit{linear}
second-order recursion
\begin{equation}
  \cphi_n(s)-(s+n+\Lambda) \cphi_{n+1}(s)+\Lambda (n+1) \cphi_{n+2}(s) = 0. 
  \label{equ:recphi}
\end{equation}
The task of deriving solutions $\phi_n$ from $\phi_N$ is now reduced to solving the recursion
(\ref{equ:recphi}). Because the latter has non-constant coefficients this remains a challenging 
problem. A systematic approach consists in utilizing the $\zcal$ transform
\begin{eqnarray}
  \label{equ:zdef} 
  \zcal\{ f_n \} &=& \sum_{n=0}^{\infty} f_n z^{-n},  \nonumber \\
  \zcal^{-1}\{ f(z) \} &=& \oint \frac{\rmd z}{2 \pi i} \, f(z) \, z^{n-1}. 
\end{eqnarray}
%
%
The transform $f(z) = \zcal\{f_n\}$ is defined in the domain $|z| > r$ with $r$ the radius
of convergence of the sum in Eq.\ (\ref{equ:zdef}). Conversely, for inverting the transform one
integrates over some closed counter-clockwise contour that is contained in the domain
of convergence of $f(z)$.

It turns out that direct $\zcal$ transformation of the sequence $\cphi_n(s)$ is not
useful for various reasons. First, specification of $\phih_N(s)$ puts constraints on
$\cphi_N(s)$ and $\cphi_{N+1}(s)$. In $\zcal$ space these translate into conditions on
the $N$ and $(N+1)$-fold derivatives of $\cphi(s;z) = \zcal\{\cphi_n(s)\}$ with respect to
$1/z$, which is inconvenient. This problem is easily avoided by renumbering the sequence
$\cphi_n(s)$ according to $\tilde{\cphi}_n(s) = \cphi_{N+1-n}(s)$. We note that \textit{a priori}
this defines the $\tilde{\cphi}_n(s)$ only for $n=0,1,\ldots, N$. It is useful, however,
to formally extend the definition of the $\tilde{\cphi}_n(s)$ over all integers $n \geq 0$
based on Eq.\ (\ref{equ:recphi}). The second problem regards convergence of the $\zcal$ transform.
Let us consider the trivial limit case $\Lambda = 0$ where the different levels of the
hierarchy (\ref{equ:MCTdef}) decouple. The recursion (\ref{equ:recphi}) now reduces to a
first-order one, yielding 
\begin{equation}
  \left. \tilde{\cphi}_n(s) \right|_{\Lambda = 0} = \prod_{k=0}^{n-1} (s+N-k). 
  \label{equ:lam0}
\end{equation}
As is conventionally done, we define empty products to give unity so that $\tilde{\cphi}_0(s) 
= \cphi_{N+1}(s) = 1$. The $\zcal$ transform (\ref{equ:zdef}) of Eq.\ (\ref{equ:lam0}) generally
diverges for all $z \in \mathbb{C}$; exceptions occur when $s$ is an integer. Therefore,
in order to use the $\zcal$ transform for analyzing Eq.\ (\ref{equ:recphi}) we have to normalize the
$\tilde{\cphi}_n(s)$ appropriately. We introduce the sequence $\psi_n(s) = \tilde{\cphi}_n(s)/n!$
or
\begin{equation}
        \psi_n(s) = \frac{\cphi_{N+1-n}(s)}{n!}. 
        \label{equ:psidef}
\end{equation}
At $\Lambda = 0$, where $\tilde{\cphi}_n(s)$ is given by Eq.\ (\ref{equ:lam0}), the $\zcal$ transform
$\psi(s;z) = \zcal\{\psi_n(s)\}$ converges for $|z| > 1$. Via a Taylor expansion in $1/z$ one shows
that in fact
\begin{equation}
  \left. \psi(s;z) \right|_{\Lambda = 0} = \left( 1 + \frac{1}{z} \right)^{s+N}. 
  \label{equ:psilam0}
\end{equation}
Expression (\ref{equ:psilam0}) also gives an analytic continuation of $\psi(s;z)$ into the
region $|z| \leq 1$. For general $s \in \mathbb{C}$ there is a branch-cut singularity in
$\psi(s;z)$ in the $z$-plane over $z \in [-1,0]$.

We now use the sequence $\psi_n(s)$ for analyzing the recursion (\ref{equ:recphi}) in the
non-trivial case $\Lambda > 0$. Expressing $\cphi_n(s)$ in terms of $\psi_n(s)$ via
Eq.\ (\ref{equ:psidef}) gives
\begin{eqnarray}
\lefteqn{  (n+1)(n+2)\psi_{n+2}(s) } \nonumber \\
&&  -  (s+\Lambda+N-n-1) (n+1) \psi_{n+1}(s) \nonumber \\
&&  +  \Lambda (N-n) \psi_n(s) = 0. 
  \label{equ:recpsi}
\end{eqnarray}
%
%
By taking the $\zcal$ transform of the latter equation, which we expect to exist for
$\Lambda > 0$, and using standard identities like $\zcal\{ n f_n \} = -z \, \partial_z \zcal\{ f_n \}$
we obtain
\begin{eqnarray}
\lefteqn{  (1+w) \partial_w^2 \psi(s;z) + \Lambda N \psi(s;z)  } \nonumber \\
&& - [s+N-1+\Lambda(1+w)] \partial_w \psi(s;z)  = 0, 
  \label{equ:psidiff}
\end{eqnarray}
%
%
where $w = 1/z$. This is a linear but non-trivial differential equation for $\psi(s;z)$. We know
from the foregoing discussion that at $\Lambda = 0$ its solution is given by Eq.\ (\ref{equ:psilam0});
this is easily verified by integration of Eq.\ (\ref{equ:psidiff}). To account for general
$\Lambda > 0$ we consider a product-ansatz of the form
\begin{equation}
  \psi(s;z) = \left(1+\frac{1}{z}\right)^{s+N} f(s;\xi) 
  \quad \mbox{where} \quad 
  \xi = \Lambda \left(1 + \frac{1}{z} \right). 
  \label{equ:subst} 
\end{equation}
The change of variable from $z$ to $\xi$ absorbs the explicit $\Lambda$-dependence in Eq.\ (\ref{equ:psidiff}).
Altogether, the substitution (\ref{equ:subst}) turns Eq.\ (\ref{equ:psidiff}) into
\begin{equation}
  \xi \partial_\xi^2 f(s;\xi) + (s+N+1-\xi) \partial_\xi f(s;\xi) - s f(s;\xi) = 0. 
  \label{equ:kummer}
\end{equation}
This equation is of the type $z \, \partial_z^2 f(z) + (b-z) \partial_z f(z) - a \, f(z) = 0$, which
is known as Kummer's differential equation. Two linearly independent solutions are given by the
functions~\cite{Mathbook}
\begin{eqnarray}
  \Phi(a,b;z) & = & \frac{1}{\Gamma(a)\Gamma(b-a)} \int_0^1 \rmd u \, u^{a-1} (1-u)^{b-a-1} \rme^{u z}, \nonumber \\ 
  \label{equ:Phiint}
  \\ 
  \Psi(a,b;z) & = & \frac{1}{\Gamma(a)} \int_0^\infty \rmd u \, u^{a-1} (1+u)^{b-a-1} \rme^{-u z}, 
  \label{equ:Psiint}
\end{eqnarray}
where $\Gamma(x)$ denotes the gamma function. The function $\Phi(a,b;z)$ is a regularized confluent
hypergeometric function, which may also be expressed as $\Phi(a,b;z) = {_1 F_1}(a,b;z)/\Gamma(b)$,
while the hypergeometric function $\Psi(a,b;z)$ is sometimes denoted $U(a,b;z)$. We note that the integral
representations (\ref{equ:Phiint}) and (\ref{equ:Psiint}) do not converge for all parameters $a,b$
and arguments $z$; however, $\Phi(a,b;z)$ and $\Psi(a,b;z)$ are well defined on $a,b,z \in \mathbb{C}$.
The regularized function $\Phi(a,b;z)$ is in fact analytic over $z \in \mathbb{C}$ while $\Psi(a,b;z)$
has a branch-cut along the negative real axis $z \in \, ]\!-\infty,0]$. We will quote further properties
of $\Phi$ and $\Psi$ as needed; see \cite{Mathbook} for a comprehensive discussion. Altogether we
have from Eqs.\ (\ref{equ:subst}) and (\ref{equ:kummer}) that the general solution of Eq.\ (\ref{equ:psidiff}) is
\begin{eqnarray}
\psi(s;z) &=& (1+w)^{s+N}  \nonumber \\
&& \times \bigg[ u_N(s) \, \Phi(s,s+N+1;\xi)   \nonumber \\ 
&& + \frac{1}{N} v_N(s) \, \Psi(s,s+N+1;\xi) \bigg], 
  \label{equ:psiuv}
\end{eqnarray}
%
with $u_N(s)$ and $v_N(s)$ arbitrary but $z$-independent functions. These account for the initial
conditions of the recursion (\ref{equ:recphi}). The factor $1/N$ in Eq.\ (\ref{equ:psiuv}) was introduced
for later convenience.

Similar to the case $\Lambda = 0$, we have from Eq.\ (\ref{equ:psiuv}) that $\psi(s;z)$ has a
branch-cut singularity in the $z$-plane over $z \in [-1,0]$ but is analytic elsewhere. We may
therefore use a circular contour containing the unit disk for inverting the $\zcal$-transform
according to Eq.\ (\ref{equ:zdef}). The functions $\cphi_n(s)$ then follow via Eq.\ (\ref{equ:psidef}) as
$\cphi_n(s) = (N+1-n)! \, \psi_{N+1-n}(s)$. From Eq.\ (\ref{equ:psiuv}) the resulting expression is
of the form $\cphi_n(s) = u_N(s) \, \mathcal{I} + (1/N) \, v_N(s) \, \mathcal{J}$ with
\begin{eqnarray}
  \mathcal{I} & = & (N+1-n)! \oint \frac{\rmd w}{2\pi i} \, w^{-(N+2-n)} (1+w)^{s+N} \nonumber \\
  && \times \Phi(s,s+N+1;\xi), 
  \label{equ:Ical} \\ 
  \mathcal{J} & = & (N+1-n)! \oint \frac{\rmd w}{2\pi i} \, w^{-(N+2-n)} (1+w)^{s+N} \nonumber \\
  && \times \Psi(s,s+N+1;\xi). 
  \label{equ:Jcal}
\end{eqnarray}
In these integrals the substitution $w = 1/z$ was made, mapping the integration contour onto a
clockwise circle contained within the unit disk; we have reverted directionality of the contour
in Eqs.\ (\ref{equ:Ical}) and (\ref{equ:Jcal}). The branch-cut
$z \in [-1,0]$ becomes $w \in \, ]\!-\infty,-1]$ in the $w$-plane. Therefore the only singularity
within the unit-disk -- and thus within the integration contour -- in both integrals is the
pole $w^{-(N+2-n)}$. For calculating the corresponding residues we use the following
representations of the hypergeometric functions,
\begin{eqnarray}
  \Phi(s,s+N+1;\xi) & = & \sum_{k=0}^\infty \frac{(s)_k}{\Gamma(s+N+k+1)} \, \frac{\xi^k}{k!}, 
  \label{equ:Phiser} \\
  \Psi(s,s+N+1;\xi) & = & \sum_{k=0}^N \binom{N}{k} \xi^{-s-k} \prod_{i=0}^{k-1} (i+s). 
  \label{equ:Psiser}
\end{eqnarray}
Equation (\ref{equ:Phiser}) follows from $\Phi(a,b;z) = {_1 F_1}(a,b;z)/\Gamma(b)$ and the
series representation~\cite{Mathbook} ${_1 F_1}(a,b;z) = \sum_{n \geq 0} [(a)_n/(b)_n] \, z^n/n!$
where $a,b,z \in \mathbb{C}$ and $(a)_n = a(a+1)\cdots (a+n-1)$. The expression (\ref{equ:Psiser}),
on the other hand, applies for $s,\xi \in \mathbb{C}$ but integer $N \geq 0$, which is
sufficient for our purposes. It is obtained from Eq.\ (\ref{equ:Psiint}) by expanding $(1+u)^N$ which
leads to integrals producing gamma functions~\cite{Mathbook}. When substituting Eqs.\ (\ref{equ:Phiser}) and
(\ref{equ:Psiser}) into Eqs.\ (\ref{equ:Ical}) and (\ref{equ:Jcal}), one readily finds the residues of the
integrands at $w = 0$. These turn out to be of the form of Eqs.\ (\ref{equ:Phiser}) and (\ref{equ:Psiser})
themselves and simply give
\begin{eqnarray}
  \mathcal{I} & = & \Phi(s,s+n;\Lambda), 
  \label{equ:Icalsol} \\ 
  \mathcal{J} & = & \frac{N!}{(n-1)!} \, \Psi(s,s+n;\Lambda). 
  \label{equ:Jcalsol} 
\end{eqnarray}
This completes inversion of the $\zcal$ transform and thus furnishes us with the general solution
$\cphi_n(s)$ of the recursion formula (\ref{equ:recphi}). Explicitly we have
\begin{equation}
  \cphi_n(s) = u_N(s) \Phi(s,s+n;\Lambda) + \frac{\Gamma(N)}{\Gamma(n)} \, v_N(s) \, \Psi(s,s+n;\Lambda). 
\end{equation}

The desired solutions $\phih_n(s)$ of our hierarchy of GMCT equations are obtained from the
$\cphi_n(s)$ using Eq.\ (\ref{equ:mapfrac}). For simplifying the resulting expression we use the
identities~\cite{Mathbook}
\begin{eqnarray}
s \, \Phi(s+1,s+n+1;\Lambda)  &=&  \Phi(s,s+n;\Lambda)  \nonumber \\
  && - n \, \Phi(s,s+n+1;\Lambda), 
  \label{equ:Phiident} \\
s \, \Psi(s+1,s+n+1;\Lambda)  &=&  - \Psi(s,s+n;\Lambda)  \nonumber \\
 && + \Psi(s,s+n+1;\Lambda). 
  \label{equ:Psiident} 
\end{eqnarray}
%
%
These are easily verified based on Eqs.\ (\ref{equ:Phiser}) and (\ref{equ:Psiser}). Our result
becomes
\begin{eqnarray}
\phih_n(s) &=& \bigg[ \Gamma(n) u_N(s) \Phi(s+1,s+n+1;\Lambda)  \nonumber \\
&& - \Gamma(N) v_N(s) \Psi(s+1,s+n+1;\Lambda) \bigg] \nonumber \\ 
&& \times \bigg[ \Gamma(n) u_N(s) \Phi(s,s+n;\Lambda)  \nonumber \\
&& + \Gamma(N) v_N(s) \Psi(s,s+n;\Lambda) \bigg]^{-1}. 
  \label{equ:gensol} 
\end{eqnarray}
%
%
It remains to work out the functions $u_N(s)$, $v_N(s)$ in terms of the initial condition $\phih_N(s)$
of the recursion (\ref{equ:contfrac}). This is done by setting $n = N$ in Eq.\ (\ref{equ:gensol}). As for
$\cphi_n(s)$ the $u_N(s)$ and $v_N(s)$ are determined only up to an overall factor. Using this freedom
we may choose
\begin{eqnarray}
  u_N(s) & = & \Psi(s+1,s+N+1;\Lambda) + \phih_N(s) \Psi(s,s+N;\Lambda), \nonumber \\
  \label{equ:uN} \\ 
  v_N(s) & = & \Phi(s+1,s+N+1;\Lambda) - \phih_N(s) \Phi(s,s+N;\Lambda). \nonumber \\
  \label{equ:vN}
\end{eqnarray}
Then, for $n = N$, Eq.\ (\ref{equ:gensol}) obviously reduces to $\phih_N(s)$. Using identities
similar to Eqs.\ (\ref{equ:Phiident}) and (\ref{equ:Psiident}), one also verifies that Eq.\ (\ref{equ:gensol}) indeed
satisfies the recursion formula (\ref{equ:contfrac}), as it should. Therefore Eqs.\ (\ref{equ:gensol})-(\ref{equ:vN})
constitute the general solution of the Mayer-Miyazaki-Reichman model, Eq.\ (\ref{equ:MCTdef}), expressing
all functions $\phih_n(s)$ with $1 \leq n \leq N$ in terms of $\phih_N(s)$. Remarkably it turns out
that rearranging Eq.\ (\ref{equ:gensol}) for $\phih_N(s)$ effectively amounts to exchanging
$n \leftrightarrow N$. This symmetry becomes obvious when expressing Eq.\ (\ref{equ:gensol}) in the form
\begin{equation}
  \Omega_n(s) = \Omega_N(s) 
  \quad \mbox{with} \quad 
  \Omega_n(s) = \Gamma(n) \, \frac{v_n(s)}{u_n(s)}. 
  \label{equ:omega} 
\end{equation}
As a consequence we may drop the restriction $n \leq N$ and Eq.\ (\ref{equ:gensol}) in fact applies for
all $n,N$; the solutions at any two levels of the hierarchy are connected via
Eq.\ (\ref{equ:gensol}). Furthermore $\Omega_n(s) = \Omega_N(s)$ for all $n,N$ implies that the expression
$\Omega_n$ itself is $n$-independent. We abbreviate $\Omega(s) \equiv \Omega_n(s)$, which is an invariant
of the recursion (\ref{equ:contfrac}) and thus of our hierarchy of GMCT equations [Eq.\ (\ref{equ:MCTdef})].

\subsection{Infinite-order solution}
\label{sec:inf}

The general solution of the Mayer-Miyazaki-Reichman model  
derived in the previous
section establishes an explicit connection between the functions $\phih_n(s)$ on all levels $n$. These
are only determined completely, however, once the closure condition (\ref{equ:cdef}) is taken into
account. Let us now analyze the behavior of the hierarchy (\ref{equ:MCTdef}) in the limit
where the closure level $N$ is taken to infinity. Here the invariant $\Omega(s)$ introduced 
in Eq.\ (\ref{equ:omega}) plays a key role. Substituting
Eqs.\ (\ref{equ:uN}) and (\ref{equ:vN}) for $u_N(s)$ and $v_N(s)$ we may express the invariant $\Omega(s)$
in terms of the closure function $\phi_N(t) = \ccal(\{\phi_n(t)\},t)$ as
\begin{eqnarray}
\lefteqn{  \Omega(s) = \Gamma(N)   } \nonumber \\
&&\times \frac{\Phi(s+1,s+N+1;\Lambda) - \phih_N(s) \Phi(s,s+N;\Lambda)} 
         {\Psi(s+1,s+N+1;\Lambda) + \phih_N(s) \Psi(s,s+N;\Lambda)}. \nonumber  \\
  \label{equ:omegaN}
\end{eqnarray}
%
%
This, in turn, determines the functions $\phih_n(s)$ on all levels $1 \leq n \leq N$ of the
hierarchy. According to Eq.\ (\ref{equ:omega}) we replace $N$ by $n$ in the latter equation and
rearrange for $\phih_n(s)$ to obtain the explicit representation
\begin{eqnarray}
\phih_n(s) &=& \bigg[ \Gamma(n) \Phi(s+1,s+n+1;\Lambda)   \nonumber \\
&& - \Omega(s) \Psi(s+1,s+n+1;\Lambda) \bigg] \nonumber \\
&& \times \bigg[ \Gamma(n) \Phi(s,s+n;\Lambda)  \nonumber \\
&& + \Omega(s) \Psi(s,s+n;\Lambda) \bigg]^{-1},
  \label{equ:phiomega}
\end{eqnarray}
%
%
where all information about the type and level of closure are absorbed in $\Omega(s)$. Therefore
solutions $\phih_n(s)$ on finite levels $n \geq 1$ of an infinite hierarchy are of the form of 
Eq.\ (\ref{equ:phiomega}) if the $N \to \infty$ limit of $\Omega(s)$ [Eq.\ (\ref{equ:omegaN})]
exists. We show in the following that a sufficient condition for this is boundedness of the sequence
of closure functions $\phi_N(t)$ over $t \geq 0$ and also derive the actual form of the solutions.

Let $s \in \mathbb{C}$ arbitrary but fixed with $\rmRe(s) > \epsilon$ for some $\epsilon > 0$.
From the triangular inequality one easily shows that boundedness of $\phi_N(t)$ over $t \geq 0$
implies that $|\phih_N(s)|$ is in turn bounded in the right half-plane $\rmRe(s) > \epsilon$. This
is all we need to know about $\phih_N(s)$ for taking the $N \to \infty$ limit in $\Omega(s)$.
To see this we rewrite (\ref{equ:omegaN}) in the form
\begin{equation}
  \Omega(s) = \lim_{N\to\infty} \frac{\Gamma(N)}{\Gamma(s+N)} \, \frac{{_1 F_1}(s,s+N;\Lambda)}{\Psi(s,s+N;\Lambda)} \, 
  \frac{X_N(s)}{Y_N(s)}, 
  \label{equ:omegaXY}
\end{equation}
where
\begin{eqnarray}
  X_N(s) & = & \frac{1}{s+N} \, \frac{{_1 F_1}(s+1,s+N+1;\Lambda)}{{_1 F_1}(s,s+N;\Lambda)} - \phi_N(s), \nonumber \\ 
  \label{equ:X_N} \\ 
  Y_N(s) & = & \frac{\Psi(s+1,s+N+1;\Lambda)}{\Psi(s,s+N;\Lambda)} + \phi_N(s). 
  \label{equ:Y_N} 
\end{eqnarray}
We have substituted $\Phi(a,b;z) = {_1 F_1}(a,b;z)/\Gamma(b)$ to simplify the discussion of the
$N \to \infty$ limit. Indeed, from the series expansion~\cite{Mathbook}
\begin{equation}
  {_1 F_1}(s,s+N;\Lambda) = \sum_{k=0}^\infty \frac{(s)_k}{(s+N)_k} \, \frac{\Lambda^k}{k!} 
  = 1 + \frac{s}{s+N} \, \Lambda + \ldots, 
  \label{equ:1F1}
\end{equation}
and convergence of the series in the right half-plane one has ${_1 F_1}(s,s+N;\Lambda) \to 1$
as $N \to \infty$ since each but the first term in (\ref{equ:1F1}) vanishes. Obviously the same
is true for ${_1 F_1}(s+1,s+N+1;\Lambda)$ which only differs in $s$ being replaced by $s+1$.
The scaling of $\Psi(s,s+N;\Lambda)$ is not so obvious. From the integral representation
(\ref{equ:Psiint}), which is convergent for $s$ in the right half-plane, we have
\begin{equation}
  \Psi(s,s+N;\Lambda) = \frac{1}{\Gamma(s)} \int_0^\infty \rmd u \, u^{s-1} (1+u)^{N-1} \, \rme^{-\Lambda u}. 
  \label{equ:PsisN}
\end{equation}
It is useful to notice that with increasing $N$ the function $(1+u)^{N-1} \, \rme^{-\Lambda u}$
develops a growing peak in $u$. A change of the integration variable $u = [(N-1)/\Lambda] v - 1$
and some rearranging produces
\begin{eqnarray}
\lefteqn{  \Psi(s,s+N;\Lambda) = \frac{\Gamma(N)}{\Gamma(s)} \, \frac{\rme^\Lambda}{\Lambda^N} 
  \left( \frac{N-1}{\Lambda} \right)^{s-1} } \nonumber \\
&& \times \int_{\frac{\Lambda}{N-1}}^\infty \rmd v \, 
  \left( v - \frac{\Lambda}{N-1} \right)^{s-1} \delta_N(v), 
  \label{equ:Psidelta} 
\end{eqnarray}
%
%
with
\begin{equation}
  \delta_N(v) = \frac{(N-1)^N}{\Gamma(N)} \, v^{N-1} \rme^{-(N-1)v}. 
\end{equation}
One shows that $\delta_N(v) \geq 0$ for $v \geq 0$, also $\int_0^\infty \rmd v \, \delta_N(v) = 1$
and, for all $v \geq 0$ except $v = 1$, $\delta_N(v) \to 0$ as $N \to \infty$. Consequently $\delta_N(v)$
is a representation of the Dirac $\delta$-distribution $\delta(v-1)$. In the limit $N \to \infty$ the
integral in Eq.\ (\ref{equ:Psidelta}) becomes $\int_0^\infty \rmd v \, v^{s-1} \delta(v-1) = 1$. Therefore
we have asymptotically
\begin{equation}
  \Psi(s,s+N;\Lambda) \sim \frac{\Gamma(N)}{\Gamma(s)} \, \frac{\rme^\Lambda}{\Lambda^N} 
  \left( \frac{N}{\Lambda} \right)^{s-1}. 
  \label{equ:Psiscal}
\end{equation}
With the above in mind the $N \to \infty$ limit of $\Omega(s)$ is easily obtained: in $X_N(s)$, 
Eq.\ (\ref{equ:X_N}), the first term vanishes due to the $1/(s+N)$ factor. Hence $|X_N(s)|$ is
bounded because $|\phih_N(s)|$ is. The first term in $Y_N(s)$, see Eq.\ (\ref{equ:Y_N}), on the
other hand, diverges like $N/(s \Lambda)$ according to Eq.\ (\ref{equ:Psiscal}). This means that
$X_N(s)/Y_N(s)$ vanishes in the right half-plane $\rmRe(s) > \epsilon$ for any bounded
$\phih_N(s)$. The remaining factor in Eq.\ (\ref{equ:omegaXY}) also vanishes on its own so
that the invariant $\Omega(s) \to 0$ for $N \to \infty$. Consequently, from Eq.\ (\ref{equ:phiomega})
the explicit solution of the infinite hierarchy is
\begin{equation}
  \phih_n(s) = \frac{\Phi(s+1,s+n+1;\Lambda)}{\Phi(s,s+n;\Lambda)}. 
  \label{equ:phihinf}
\end{equation}

This is a highly non-trivial result. As the closure level $N$ is taken to infinity, the functions
$\phih_n(s)$ on all finite levels $n \geq 1$ become independent of the particular closure used.
Intuitively we can understand this in the following
way: at large $N$ and for sufficiently small times $t$ the significant terms in Eq.\ 
(\ref{equ:MCTdef}) at level $N-1$ are $\partial_t \phi_{N-1}(t) + (N-1) \, \phi_{N-1}(t) 
\approx 0$. Therefore $\phi_{N-1}(t) \approx \rme^{-(N-1)t}$ initially drops rapidly
with $t$. The closure function $\phi_N(t)$ only becomes relevant in Eq.\ (\ref{equ:MCTdef})
once $\phi_{N-1}(t)$ has dropped to a tiny fraction of its initial value
$\phi_{N-1}(0) = 1$. Our result (\ref{equ:phihinf}) shows that as we
descend to lower levels $N-2,N-3,\ldots$ contributions of $\phi_N(t)$ are washed
out. In the limit $N \to \infty$, where we have to descend through infinitely many
levels of the hierarchy in order to reach a finite level $n$, residual contributions
of $\phi_N(t)$ disappear. More precisely, Eq.\ (\ref{equ:phihinf}) is an \textit{attractor} for
downward-recursion of Eq.\ (\ref{equ:contfrac}). The attraction basin comprises at least
all bounded closure functions.

\begin{figure}
        \begin{center}
    \epsfig{file=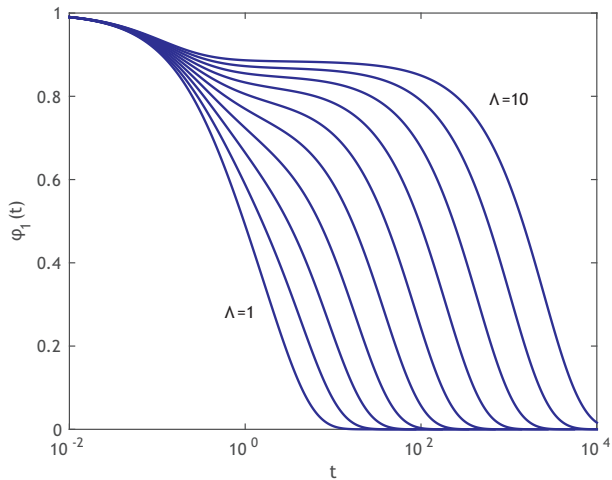,width=8cm,clip}
  \end{center}
  \caption{
  \label{fig:Phi1t} The solutions $\phi_1(t)$ of the infinite Mayer-Miyazaki-Reichman 
  hierarchy 
  for $\Lambda = 1,2,\ldots 10$ from
  left to right, respectively. The function $\phi_1(t)$ shows beta-relaxation
  for $t = \mathcal{O}(1)$, has a plateau in the range $1 \ll t \ll \taua$, and
  eventually decays exponentially in $t/\taua$.
  }
\end{figure}

Let us now
investigate the predictions of Eq.\ (\ref{equ:phihinf}). To do so we have to
invert the Laplace transform of $\phih_n(s)$. We use that the regularized
hypergeometric functions $\Phi$ appearing in the numerator and denominator of Eq.\
(\ref{equ:phihinf}) are analytic in $s$. Therefore the only singularities of
$\phih_n(s)$ are poles located at the zeros $\{s_k\}$ of the denominator
in Eq.\ (\ref{equ:phihinf}), i.e.,
\begin{equation}
  \Phi(s_k,s_k+n,\Lambda)=0. 
  \label{equ:sk}
\end{equation}
It turns out that all $s_k$ lie on the negative real axis. Consider, for a moment,
the trivial case $\Lambda = 0$. Then $\Phi(s_k,s_k+n,0) = 1/\Gamma(s_k+n)=0$ has
solutions $s_k = -n-k$ with $k=0,1,2,\ldots$. Numerical analysis of Eq.\ (\ref{equ:sk})
shows that for $\Lambda > 0$ the $s_k$ shift toward the origin, however,
we always have that $s_0 < 0$ and $s_{k+1} \leq s_k - 1$. Thus, the inverse
Laplace transform of Eq.\ (\ref{equ:phihinf}) is
\begin{equation}
  \phi_n(t) = \sum_{k=0}^\infty r_k \, \rme^{s_k t} 
  \quad \mbox{with} \quad 
  r_k = \mathrm{Res}\left( \phih_n(s), s \to s_k \right).  
  \label{equ:phint}
\end{equation}
The roots $\{s_k\}$ and residues $\{r_k\}$ of course depend on $\Lambda$ and $n$; however,
we omit these arguments for brevity. Because the difference between successive $s_k$'s is
at least one there is only a \textit{single} slow mode $r_0 \rme^{s_0 t}$ in the system. The
remaining sum $\sum_{k \geq 1} r_k \rme^{s_k t}$ in Eq.\ (\ref{equ:phint}) is $\ocal(\rme^{-t})$ and
only contributes to fast beta-relaxation processes. Consequently there is exponential
alpha-relaxation in the infinite hierarchy with alpha-relaxation time
\begin{equation}
  \taua=-\frac{1}{s_0}. 
  \label{equ:taua}
\end{equation}
Numerical evaluation of Eqs.\ (\ref{equ:sk}) and (\ref{equ:phint}) is straightforward.
In particular it turns out that the sum in Eq.\ (\ref{equ:phint}) converges rapidly, but with the
number of relevant terms (for beta-relaxation) increasing with $\Lambda$. Plots of the
time dependence of $\phi_1(t)$ in the infinite hierarchy and for various $\Lambda$ are shown
in Fig.~\ref{fig:Phi1t}.

%

In fact, analysis of Eq.\ (\ref{equ:sk}) allows us to deduce the scaling of the
alpha-relaxation time $\taua$ of $\phi_1(t)$ for $\Lambda \gg 1$. From Eq.\ (\ref{equ:Phiint})
we have $\Phi(s,s+1;\Lambda) = [1/\Gamma(s)] \int_0^1 \rmd u \, u^{s-1} \, \rme^{\Lambda u}$.
This integral representation converges for $\rmRe(s) > 0$, however, the alpha-relaxation
time is determined by the root $-1 \leq s_0 < 0$. We use that via integration by parts
$\Phi(s,s+1;\Lambda) = [1/\Gamma(s+1)] [ \rme^\Lambda - \Lambda \int_0^1 \rmd u \, u^s \, 
\rme^{\Lambda u}]$, which provides an analytic continuation down to $\rmRe(s) > -1$.
This expression vanishes for a $-1 \leq s_0 < 0$ satisfying
$\Lambda \int_0^1 \rmd u \, u^{s_0} \, \rme^{\Lambda u} = \rme^\Lambda$.
Changing the integration
variable $v = \Lambda (1-u)$ turns this into $\int_0^\Lambda \rmd v \, (1-v/\Lambda)^{s_0} 
\rme^{-v} = 1$ which is suitable for a $\Lambda \to \infty$ asymptotic expansion: due to
the $\rme^{-v}$ factor the integral
is dominated by contributions from finite $v$. We may thus Taylor expand $(1-v/\Lambda)^{s_0}$
in $v/\Lambda$ and, by truncating at first order, obtain via Eq.\ (\ref{equ:taua})
\begin{equation}
  \taua \sim \frac{\rme^\Lambda}{\Lambda} 
  \quad \mbox{for} \quad 
  \Lambda \to \infty. 
  \label{equ:arr}
\end{equation}
Clearly there is no divergence of $\taua$ at any finite $\Lambda$ in the
infinite-order solution.  In cases where $\Lambda \propto 1/T$, for instance,
the result (\ref{equ:arr}) essentially predicts Arrhenius behavior of the
alpha-relaxation time. The emergence of such a relaxation time scale that depends
exponentially on the coupling, which is a hallmark of activated behavior,
is remarkable. This is to be contrasted with the results of standard
MCT or, as shown below, any level of ''mean-field closure",
which always predicts a relaxation time that diverges as a power law at a
finite value of the control parameter. Hence, the infinite-order construct of
GMCT gives rise to a fundamentally new relaxation behavior. This concludes our
explicit derivation of the infinite-order solution of the
Mayer-Miyazaki-Reichman model, the result of which will serve as a benchmark
for the finite-order closures discussed in the following subsections.


\subsection{Mean-field closures at finite order}
\label{sec:mfn}
We will now consider solutions of the Mayer-Miyazaki-Reichman model for specific closure conditions
[Eq.\ (\ref{equ:cdef})] at \textit{finite} closure levels $N$. 
We will first look at so-called ``mean-field closures" of the form
\begin{equation}
  \phi_N(t) = \ccal(\{\phi_n(t)\},t)=\phi_{n_1}^{m_1}(t) \phi_{n_2}^{m_2}(t) \cdots \phi_{n_k}^{m_k}(t). 
  \label{equ:cmfn} 
\end{equation}
The simplest example of this type is the classical factorization $\phi_2(t) = \phi_1^2(t)$, i.e.,
the Leutheusser model~\cite{leutheusser:84}; others would be $\phi_3(t) = \phi_1^3(t)$ or
$\phi_3(t) = \phi_1(t) \phi_2(t)$ and so on. We introduce the notation MF-$N$ for mean-field
closures at level $N$, or more specifically MF-$N({n_i}^{m_i})$ to precisely reflect
Eq.\ (\ref{equ:cmfn}). The examples listed above would correspond to MF-$2(1^2)$,
MF-$3(1^3)$ and MF-$3(1^1 2^1)$, respectively. Any mean-field closure should satisfy
\begin{equation}
  N = n_1 m_1 + n_2 m_2 + \ldots + n_k m_k, 
  \label{equ:Nmatch}
\end{equation}
where $1 \leq n_1 < n_2 < \ldots n_k < N$ and $m_1, m_2, \ldots m_k \in \mathbb{N}$. This constraint
expresses that the total order of correlations on the right hand side of Eq.\ (\ref{equ:cmfn}) is required
to match with $\phi_N(t)$. From Eq.\ (\ref{equ:Nmatch}) the number of different mean-field closures at
level $N$ grows rapidly with $N$ -- namely as the number of integer partitions of $N$.
The first few numbers for $N = 2,\ldots,6$ are $1,2,4,6,10$.

\begin{figure}
  \epsfig{file=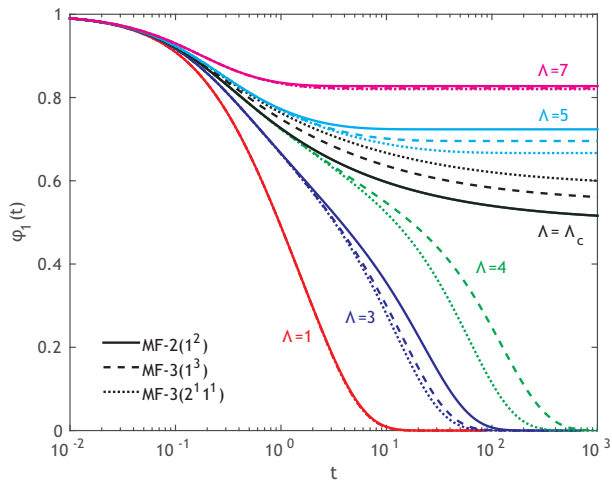,width=8cm,clip}
  \caption{\label{fig:MF3} Solutions $\phi_1(t)$ for the
Mayer-Miyazaki-Reichman model under various low-order MF-$N$ closures.
The critical points for the different closures are $\Lambda_c=4$ for MF-2($1^2$),
$\Lambda_c \approx 4.4922$ for MF-2($1^3$), and $\Lambda_c \approx 4.8284$ for MF-2($2^1 1^1$).
}
\end{figure}

For mean-field closures (\ref{equ:cmfn}) it is not possible to derive the exact time dependence
of solutions $\{\phi_n(t)\}$. On the one hand, we have a Laplace transformed representation of the general
solution (\ref{equ:gensol}) which cannot be inverted explicitly, while on the other, Laplace transformation
of the closure (\ref{equ:cmfn}) yields multiple complex convolution integrals and is not useful either.
We thus limit the analysis in this section to the identification of features of the solutions $\{\phi_n(t)\}$,
in particular the appearance of a dynamical transition and scaling near the transition. The discussion
will be along the lines of usual schematic MCT~\cite{leutheusser:84,bengtzelius:84}. Data for the time
dependence of $\{ \phi_n(t) \}$ is obtained numerically. We use an adaptation of the algorithm introduced
in Ref.\ \cite{fuchs:91} which is directly based on Eqs.\ (\ref{equ:MCTdef}) and (\ref{equ:cdef}). Representative
examples of solutions $\phi_1(t)$ are shown in Fig.~\ref{fig:MF3} for several low-order MF closures.

It may be seen that for all mean-field closures considered in Fig.~\ref{fig:MF3}, the dynamics slows
down considerably with increasing $\Lambda$, and for sufficiently high $\Lambda$ undergoes a 
discontinuous transition that is reminiscent of the transition predicted by Leutheusser's MF-$2(1^2)$ model.
Indeed, the numerical data in Fig.~\ref{fig:MF3} suggests that
any mean-field closure has a critical parameter $\Lambda_c$ at which a dynamical transition occurs:
for $\Lambda < \Lambda_c$ all $\phi_n(t)$ vanish at long times $t \to \infty$, 
whereas for $\Lambda \geq \Lambda_c$ the functions $\phi_n(t)$ approach a plateau $0 < q_n < 1$,
\begin{equation} 
  q_n = \lim_{t \to \infty} \phi_n(t). 
  \label{equ:qndef}
\end{equation}
Note that $q_n$ may serve as an order parameter for the transition, and is 
sometimes referred to as the nonergodicity parameter. 
A bifurcation analysis of the fixed-point equation for $q_n$, which we derive in the following,
will allow us to determine the critical parameter $\Lambda_c$. First we utilize the general solution
(\ref{equ:gensol}) to express all $\phih_n(s)$ in terms of $\phih_1(s)$. It is convenient to
introduce functions $A_n(s), B_n(s), C_n(s)$ and $D_n(s)$ such that
\begin{equation}
  \phih_n(s) = - \frac{C_n(s) - D_n(s) \, \phih_1(s)}{A_n(s) - B_n(s) \, s \, \phih_1(s)}. 
  \label{equ:phiAD}
\end{equation}
Substituting Eqs.\ (\ref{equ:uN}) and (\ref{equ:vN}) into Eq.\ (\ref{equ:gensol}) and comparing coefficients
with Eq.\ (\ref{equ:phiAD}) defines $A_n(s), \ldots D_n(s)$. Explicit expressions are given in
Eqs.\ (\ref{equ:Ans})-(\ref{equ:Dns}) in the Appendix. From Eq.\ (\ref{equ:phiAD}) one obtains the $q_n$'s in
terms of $q_1$ using the general identity for Laplace transforms
\begin{equation}
  \lim_{t \to \infty} \phi_n(t) = \lim_{s \to 0} s \, \phih_n(s). 
  \label{equ:lim}
\end{equation}
We discuss in the Appendix that $A_n(s),\ldots D_n(s)$ are entire functions of $s$ (this is somewhat
non-trivial for $B_n(s)$ [Eq.\ (\ref{equ:Bns})], which contains a factor $s^{-1}$ that compensates
for the explicit $s$ introduced in the denominator of Eq.\ (\ref{equ:phiAD}) above). Thus, the $s \to 0$ limit
of $A_n(s), \ldots D_n(s)$, which we denote $A_n, \ldots D_n$, exists. From the discussion in the
Appendix
\begin{eqnarray}
  A_n & = & \frac{1}{\Lambda} \, \rme^\Lambda, 
  \label{equ:An} \\ 
  B_n & = & \frac{1}{\Lambda} \, \rme^\Lambda \sum_{k=0}^{n-2} \frac{k!}{\Lambda^k}, 
  \label{equ:Bn} \\ 
  C_n & = & \frac{(n-1)!}{\Lambda^n} \, \rme^\Lambda \sum_{k=0}^{n-2} \frac{\Lambda^k}{(k+1)!}, 
  \label{equ:Cn} \\
  D_n & = & \frac{(n-1)!}{\Lambda^n} \, \rme^\Lambda. 
  \label{equ:Dn} 
\end{eqnarray}
When expanding Eq.\ (\ref{equ:phiAD}) by $s$, and considering the explicit factor $s$ introduced in
the denominator, the $\phih_n(s)$ only appear in the combination $s \, \phih_n(s)$. According
to Eqs.\ (\ref{equ:qndef}) and (\ref{equ:lim}) this reduces to $q_n$ for $s \to 0$ and therefore
Eq.\ (\ref{equ:phiAD}) produces
\begin{equation}
  q_n = \frac{D_n \, q_1}{A_n - B_n \, q_1}. 
  \label{equ:qnq1}
\end{equation}
This specific expression for the long time limits $q_n$ of the functions $\phi_n(t)$ is a
consequence of our hierarchy of GMCT equations (\ref{equ:MCTdef}). Combining Eq.\ (\ref{equ:qnq1})
with the mean-field closure (\ref{equ:cmfn}), which for $t \to \infty$ simply reduces to
\begin{equation}
  q_N = \prod_{i = 1}^k q_{n_i}^{m_i}, 
  \label{equ:qN}
\end{equation}
yields the fixed-point equation for $q_1$.
Via substitution of Eqs.\ (\ref{equ:qnq1}) and (\ref{equ:An})-(\ref{equ:Dn}) and rearranging,
keeping in mind Eq.\ (\ref{equ:Nmatch}), we rewrite Eq.\ (\ref{equ:qN}) in the form $P(q_1) = 0$ with
\begin{eqnarray}
P(q) &=& \prod_{i = 1}^k \left[ \frac{1}{\Lambda (n_i-1)!} \left( \frac{1}{q} - 
  \sum_{n=0}^{n_i -2} \frac{n!}{\Lambda^n} \right) \right]^{m_i}  \nonumber \\
&& - \frac{1}{\Lambda (N-1)!} \left( \frac{1}{q} - \sum_{n=0}^{N-2} \frac{n!}{\Lambda^n} \right).  
  \label{equ:FP}
\end{eqnarray}
%

\begin{figure}
  \epsfig{file=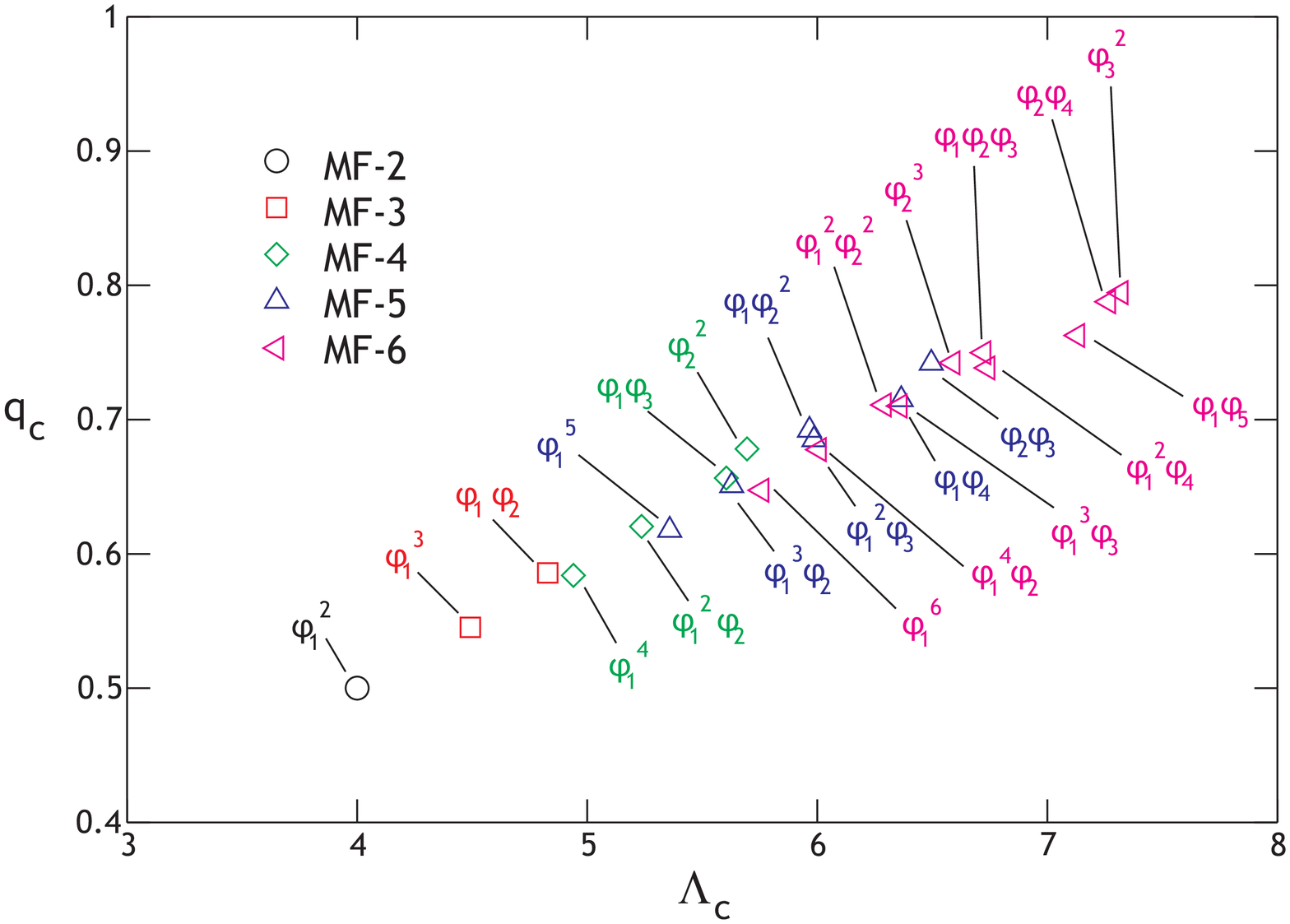,width=0.45\textwidth,clip}
  \caption{\label{fig:mfn} Plateau height $q_c$ of $\phi_1(t)$ at criticality 
versus the predicted critical point $\Lambda_c$
for various MF-$N$ closures of the Mayer-Miyazaki-Reichman hierarchy.
}
\end{figure}

Equation (\ref{equ:FP}), which is a polynomial of degree $\sum_i m_i$ in
$1/q$, is our key to locate the dynamical transition for generic MF-$N$ mean-field 
closures. Such transitions are characterized by the appearance of a real root
$q_c \in \, ]0,1[$ as $\Lambda$ is raised to the critical value $\Lambda_c$.
We have analyzed Eq.\ (\ref{equ:FP}) numerically for all MF-$N$ closures up to level
$N=6$. Figure~\ref{fig:mfn} shows corresponding numerical data.
For each closure there is a $\Lambda_c$ such that for $\Lambda \in \, ]0,\Lambda_c[$ Eq.\ 
(\ref{equ:FP}) has no real roots $q \in \, ]0,1[$. Consequently, Eqs.\ (\ref{equ:qnq1}) and (\ref{equ:qN})
only admit the trivial solution $q_1 = 0$ and the $\phi_n(t)$ vanish at long times. At
$\Lambda = \Lambda_c$ a pair of complex conjugate roots merges into the quadratic root $q = q_c$,
however, and the $\phi_n(t)$ do not relax anymore but approach the plateau $q_c$ as $t \to \infty$.
Plots of critical solutions $\phi_1(t)$ are contained in Fig.~\ref{fig:MF3} above. For $\Lambda 
> \Lambda_c$ the pair of real roots of Eq.\ (\ref{equ:FP}) separates: one decreases towards $q \to 0$
while the other increases $q \to 1$ as $\Lambda$ grows $\Lambda \to \infty$. The larger of the
two is the physically relevant root~\cite{bengtzelius:84} and defines the plateau height in $\phi_1(t)$
at $\Lambda > \Lambda_c$.
Note that we find the general trend that closing at a higher level $N$ increases $\Lambda_c$ and $q_c$.
Moreover, among the various closures possible at any given level $N$, the MF-$N(1^N)$ closure
always seems to yield the lowest $\Lambda_c$. 

In order to investigate in more detail the dependence of $\Lambda_c$ on the closure level $N$
as $N$ becomes large, we specialize to MF-$N(1^N)$ closures. In that case Eq.\ 
(\ref{equ:FP}) is particularly simple. Setting $k=1, n_1=1, m_1=N$ and some rearranging allows
us to rewrite Eq.\ (\ref{equ:FP}) in the form
\begin{equation}
  P(q_1) = 0 
  \quad \mbox{with} \quad 
  P(q) = q^N \sum_{n=0}^{N-2} \frac{n!}{\Lambda^n} - q^{N-1} + \frac{(N-1)!}{\Lambda^{N-1}}. 
  \label{equ:FP1N}
\end{equation}
Clearly $P(0) > 0$ and $P(q)$ has a single local extremum, which is in fact a minimum,
in the range $q > 0$ for any $N \geq 2$ and $\Lambda > 0$. These facts and the actual form of
Eq.\ (\ref{equ:FP1N}) make it easy to derive an equation for $\Lambda_c$: the polynomial $\partial_q 
P(q)|_{q_x} = 0$ can be solved for the location $q_x(\Lambda)$ of the extremum. At criticality
we then have $P\big(q_x(\Lambda_c)\big) = 0$, or explicitly
\begin{equation}
  \frac{N}{N-1} \sum_{n=0}^{N-2} \frac{n!}{\Lambda_c^n} = \frac{\Lambda_c}{\sqrt[N-1]{N!}}. 
  \label{equ:lc}
\end{equation}
From this, one extracts the leading asymptotic form $\Lambda_c \sim N/\rme$ at large $N$, using
that $\sqrt[N]{N!}/N \to 1/\rme$ for $N \to \infty$ and dominance of the first term in the sum
in Eq.\ (\ref{equ:lc}). The plateau height $q_c = q_x(\Lambda_c)$ at the critical point $\Lambda = 
\Lambda_c$ is found to be $q_c = \sqrt[N-1]{N!}/\Lambda_c$ and approaches unity $q_c \to 1$ as
$N \to \infty$. Hence, in MF-$N(1^N)$ closures the critical parameter $\Lambda_c$ scales \textit{linearly}
with the level $N$ of closure of the hierarchy. For $N \to \infty$ we have $\Lambda_c \to \infty$
and the mode-coupling transition disappears completely, conform the infinite-order result of 
Eq.\ (\ref{equ:arr}).
This highlights an important and non-trivial result: closing the hierarchy at increasingly higher order
yields a critical point that converges systematically towards the exact infinite-order solution
$\Lambda_c \rightarrow \infty$. \textit{This uniform convergence thus emerges naturally
from our GMCT model, even though no small perturbation parameter is present in the theory.}

Having determined the critical points $\Lambda_c$ of generic MF-$N(n_i^{m_i})$ closures
we next focus our attention on the time dependence of the corresponding solutions $\phi_n(t)$.
We isolate the known plateau values $q_n$ via the ansatz
\begin{equation} 
  \phi_n(t) = q_n + \mu_n(t), 
  \label{equ:phimu} 
\end{equation}
consequently $\mu_n(t) \to 0$ for $t \to \infty$. Our aim in the following will be to determine
the asymptotic shape of the beta-relaxation functions $\mu_n(t)$. Using the Laplace transformed
representation $\phih_n(s) = q_n/s + \muh_n(s)$ of Eq.\ (\ref{equ:phimu}), Eq.\ (\ref{equ:phiAD})
may be rearranged in the form
\begin{equation}
  \muh_n(s) = \frac{E_n(s)+[D_n(s)+q_n B_n(s)]\muh_1(s)}{A_n(s)-B_n(s) [q_1 + s \, \muh_1(s)]}. 
  \label{equ:munmu1} 
\end{equation}
Here we have introduced the new function
\begin{equation}
  E_n(s) = -C_n(s) + [ q_1 D_n(s) - q_n A_n(s) + q_1 q_n B_n(s) ] \, s^{-1}. 
  \label{equ:Ens} 
\end{equation}
Note that in the limit $s \to 0$ the coefficient of $s^{-1}$ in Eq.\ (\ref{equ:Ens}) vanishes like
$\ocal(s)$. This follows from Taylor expansion of $A_n(s), B_n(s), D_n(s)$ and the fact that
the $q_n$ satisfy Eq.\ (\ref{equ:qnq1}). Therefore the $s^{-1}$ singularity in Eq.\ (\ref{equ:Ens}) is
lifted and $E_n(s)$ is an analytic function of $s \in \mathbb{C}$. In analogy to $B_n(s)$
we define $E_n(s)$ at $s=0$ via the limit $E_n = E_n(s \to 0)$.

As usual we complement Eq.\ (\ref{equ:munmu1}) with the mean-field closure (\ref{equ:cmfn}) to
determine the functions $\phi_n(t)$ and thus $\mu_n(t)$. Plugging our ansatz (\ref{equ:phimu}) into
Eq.\ (\ref{equ:cmfn}) produces
\begin{eqnarray} 
 q_N + \mu_N(t) &=& [q_{n_1} + \mu_{n_1}(t)]^{m_1} [q_{n_2} + \mu_{n_2}(t)]^{m_2}  \nonumber \\
&& \cdots 
  [q_{n_k} + \mu_{n_k}(t)]^{m_k}. 
  \label{equ:muNprod}
\end{eqnarray}
By expanding binomials and products and using the fact that the $q_n$ satisfy Eq.\ (\ref{equ:qN}),
we may rewrite this expression in the form
\begin{equation}
  \mu_N(t) = q_N \sum_{i=1}^k \frac{m_i}{q_{n_i}} \mu_{n_i}(t) + \bcal(t), 
  \label{equ:Btdef}
\end{equation}
where $\bcal(t)$ contains bilinear and higher order terms in $\mu_n(t)$. From Eq.\ (\ref{equ:muNprod}),
explicitly writing out $\bcal(t)$ to bilinear order amounts to
\begin{eqnarray}
  \bcal(t) &=& q_N \Bigg[ \sum_{i=1}^k \frac{m_i (m_i-1)}{2 q_{n_i}^2} \, \mu_{n_i}^2(t) 
\nonumber \\
&&  + \sum_{1 \leq i < j \leq k} \frac{m_i m_j}{q_{n_i} q_{n_j}} \, \mu_{n_i}(t) 
  \mu_{n_j}(t) \Bigg] + \ldots 
  \label{equ:Bt} 
\end{eqnarray}
%

The difficulty of solving the system of equations (\ref{equ:munmu1}), (\ref{equ:Btdef}) and
(\ref{equ:Bt}) lies in the fact that $\bcal(t)$ as given in Eq.\ (\ref{equ:Bt}) is non-linear in
the functions $\mu_n(t)$. A useful expression for the Laplace transform $\hat{\bcal}(s) = 
\lcal \{ \bcal(t) \}$ can only be derived from Eq.\ (\ref{equ:Bt}) if the $\mu_n(t)$ are known
explicitly. In order to proceed we now focus on asymptotic behavior and make the ansatz
\begin{equation}
  \mu_1(t) \sim c \, t^{-\beta} 
  \quad \mbox{for} \quad 
  t \to \infty, 
  \label{equ:mutinf}
\end{equation}
with $c > 0$ some amplitude and $\beta$ the exponent for beta-relaxation; in the mode-coupling
literature this exponent is usually denoted $a$; however, we follow here the more appropriate
notation of \cite{bouchard:96}. We shall prove in the
following that $\mu_1(t)$ indeed displays power-law relaxation for any mean-field closure
and determine the value of the exponent $\beta$. We assume that $\beta$ satisfies $\frac{1}{3} < 
\beta < \frac{1}{2}$ which will turn out to be the case. The asymptotic power-law behavior
(\ref{equ:mutinf}) implies for $\muh_1(s) = \lcal \{ \mu_1(t) \}$,
\begin{equation}
  \muh_1(s) \sim c \, \Gamma(1-\beta) \, s^{\beta-1} 
  \quad \mbox{for} \quad 
  s \to 0. 
  \label{equ:mus0}
\end{equation}
To leading order the Laplace transform $\muh_1(s)$ has power-law divergence at the origin. From
Eq.\ (\ref{equ:munmu1}) it follows that in fact all $\muh_n(s)$ are singular at $s = 0$. Using that
$A_n(s),\ldots, E_n(s)$ have finite $s \to 0$ limits and that $s \, \mu_1(s) \to 0$ for $s \to 0$
we obtain from Eq.\ (\ref{equ:munmu1})
\begin{equation}
  \muh_n(s) \sim \frac{D_n+q_n B_n}{A_n-q_1 B_n} \, \muh_1(s) = \frac{q_n}{q_1} 
  \frac{A_n}{A_n - q_1 B_n} \, \muh_1(s), 
  \label{equ:muns0} 
\end{equation}
for $s \to 0$. Here Eq.\ (\ref{equ:qnq1}) was used to simplify the prefactor of $\muh_1(s)$. Via
inversion of the Laplace transform we have in turn
\begin{equation}
  \mu_n(t) \sim \frac{q_n}{q_1} \frac{A_n}{A_n - q_1 B_n} \, \mu_1(t) 
  \quad \mbox{for} \quad 
  t \to \infty. 
  \label{equ:muntinf}
\end{equation}
Thus, all functions $\mu_n(t)$ decline asymptotically like $t^{-\beta}$. Consequently, $\bcal(t)$ has
$t^{-2\beta}$ asymptotics from bilinear contributions in Eq.\ (\ref{equ:Bt}). This, in turn, implies a
$s^{2\beta-1}$ divergence -- since $\beta < \frac{1}{2}$ is assumed -- of the Laplace transform $\hat{\bcal}(s)$
for $s \to 0$. Higher order terms in Eq.\ (\ref{equ:Bt}) are $\ocal(t^{-3\beta})$ at large $t$ which, as
we assume $\beta > \frac{1}{3}$, leads to a convergent Laplace integral (\ref{equ:Ldef}) for
$s \to 0$ and thus subdominant $\ocal(1)$ contributions to $\hat{\bcal}(s)$. Altogether,
the leading asymptotics of $\hat{\bcal}(s) = \lcal\{ \bcal(t) \}$ for $s \to 0$ are obtained
by substituting Eq.\ (\ref{equ:muntinf}) in Eq.\ (\ref{equ:Bt}) and retaining bilinear terms only.
The identity
\begin{equation}
  \frac{A_N}{A_N - q_1 B_N} = \sum_{i = 1}^k m_i \frac{A_{n_i}}{A_{n_i} - q_1 B_{n_i}}, 
  \label{equ:exid}
\end{equation}
which only applies at $\Lambda = \Lambda_c$, is useful for simplifying the result. It follows
from Eq.\ (\ref{equ:FP}) and the fact that at criticality $q_c$ is a quadratic root of $P(q)$, i.e.,
$\partial_q P(q) |_{q = q_c} = 0$. We finally rewrite the two-dimensional sum in Eq.\ (\ref{equ:Bt})
as the square of Eq.\ (\ref{equ:exid}) to find the $s \to 0$ scaling
\begin{eqnarray} 
  \hat{\bcal}(s) &\sim& c^2 \, \Gamma(1-2\beta) \, \frac{q_N}{2 q_1^2} \Bigg[ 
  \left( \frac{A_N}{A_N - q_1 B_N} \right)^2 
\nonumber \\
&& - \sum_{i=1}^k m_i 
  \left( \frac{A_{n_i}}{A_{n_i}-q_1 B_{n_i}} \right)^2 \Bigg] 
  s^{2\beta-1}. 
  \label{equ:Bs1}
\end{eqnarray}
%
%
The coefficient in the square brackets is related to $\partial_q^2 P(q) |_{q=q_c}$ and should
be non-zero. This $s \to 0$ scaling is a consequence of the power-law ansatz (\ref{equ:mutinf})
and Eqs.\ (\ref{equ:munmu1}) and (\ref{equ:Bt}).

In order for the power-law ansatz (\ref{equ:mutinf}) to apply, however, the result (\ref{equ:Bs1})
must be consistent with the scaling of $\hat{\bcal}(s)$ as determined by Eqs.\ (\ref{equ:munmu1}) and
(\ref{equ:Btdef}). From rearranging Eq.\ (\ref{equ:Btdef}), we have
\begin{equation}
  \hat{\bcal}(s) = \muh_N(s) - q_N \sum_{i=1}^k \frac{m_i}{q_{n_i}} \muh_{n_i}(s). 
  \label{equ:Bsalt}
\end{equation}
If we substitute the leading $s \to 0$ scaling (\ref{equ:muns0}) for the $\muh_n(s)$, then the right
hand side of Eq.\ (\ref{equ:Bsalt}) vanishes; this is seen from the identity (\ref{equ:exid}). Therefore,
in order to derive the leading $s \to 0$ scaling of $\hat{\bcal}(s)$ from Eq.\ (\ref{equ:Bsalt}), we need
an expansion of the $\muh_n(s)$ to first sub-dominant order. Contributions of $E_n(s)$ to $\muh_n(s)$
in Eq.\ (\ref{equ:munmu1}) are $\ocal(1)$ near $s = 0$ as $s \, \muh_1(s)$ vanishes while $A_n(s)$, \ldots
$E_n(s)$ have finite $s \to 0$ limits. Thus,
\begin{equation}
  \muh_n(s) = \frac{D_n(s)+q_n B_n(s)}{A_n(s)-B_n(s) [q_1 + s \, \muh_1(s)]} \, \muh_1(s) + \ocal(1). 
\end{equation}
Next consider the fraction in this expression: for obtaining an expansion to first subleading order we
recall that the functions $A_n(s),\ldots D_n(s)$ are analytic and therefore admit Taylor expansions at
$s=0$. We have $A_n(s) = A_n + \ocal(s)$ etc. On the other hand, the denominator contains a term
$s \, \muh_1(s)$ which is $\ocal(s^\beta)$. As $\beta < 1$ the first subdominant correction near $s=0$ originates
from the latter term alone and we get
\begin{equation}
  \muh_n(s) = \frac{D_n+q_n B_n}{A_n-q_1 B_n} \left[ 1 
  + \frac{B_n}{A_n-q_1 B_n} \, s \, \muh_1(s) 
  + \ldots \right] \, \muh_1(s). 
  \label{equ:munsub}
\end{equation}
According to Eq.\ (\ref{equ:mus0}) the functions $\muh_1(s)$ and $s \, \muh_1^2(s)$ diverge like $s^{\beta-1}$ and
$s^{2\beta-1}$ for $s \to 0$, respectively. Higher order corrections in Eq.\ (\ref{equ:munsub}), which diverge
slower than that or remain $\ocal(1)$, are irrelevant. Simplifying the prefactor in Eq.\ (\ref{equ:munsub})
as in Eq.\ (\ref{equ:muns0}) above, substitution into Eq.\ (\ref{equ:Bsalt}) and making use of the identity
(\ref{equ:exid}) to cancel the leading contribution from Eq.\ (\ref{equ:munsub}) leaves us with
\begin{eqnarray}
  \hat{\bcal}(s) &\sim& c^2 \, \Gamma^2(1-\beta) \, \frac{q_N}{q_1^2} \Bigg[ 
  \left( \frac{A_N}{A_N - q_1 B_N} \right)^2 
\nonumber \\
&& - \sum_{i=1}^k m_i 
  \left( \frac{A_{n_i}}{A_{n_i}-q_1 B_{n_i}} \right)^2
  \Bigg] s^{2\beta-1}. 
  \label{equ:Bs2}
\end{eqnarray}
%
%
Equations (\ref{equ:Bs1}) and (\ref{equ:Bs2}) produce a consistent result for the $s \to 0$ scaling
of $\hat{\bcal}(s)$. Our ansatz (\ref{equ:mutinf}) for the asymptotic form of $\mu_1(t)$ is therefore
valid. The beta-relaxation exponent $\beta$ is determined by matching the coefficients of $s^{2\beta-1}$
in (\ref{equ:Bs1}) and (\ref{equ:Bs2}). This simply requires
\begin{equation}
  2 \, \Gamma^2(1-\beta) = \Gamma(1-2\beta), 
  \label{equ:bfix}
\end{equation}
which has a numerical solution $\beta = 0.395\ldots$, satisfying our assumption $\frac{1}{3} < \beta < 
\frac{1}{2}$. This is a remarkable result because it applies for \textit{arbitrary} MF-$N$
mean-field closures. The MF-$2(1^2)$ closure corresponding to the Leutheusser model, where this
exponent was first derived \cite{leutheusser:84,bengtzelius:84}, is just a special case of our general
result. The fact that the closure-dependent factors in Eqs.\ (\ref{equ:Bs1}) and (\ref{equ:Bs2}) --
i.e.\ the ratio $q_N/q_1$ and the expression in the square brackets -- turn out to be the same is
non-trivial.

In summary, beta-relaxation in the Mayer-Miyazaki-Reichman model [Eq.\
(\ref{equ:MCTdef})], subject to an arbitrary mean-field closure [Eq.\ (\ref{equ:cmfn})],
is always of the form $\mu_1(t) \sim c \, t^{-\beta}$ at long times, with a
unique exponent $\beta$ defined by Eq.\ (\ref{equ:bfix}). The location of the
critical point $\Lambda = \Lambda_c$ at which this occurs, however, does depend
on the particular mean-field closure used. From Fig.~\ref{fig:mfn} and the
foregoing discussion, $\Lambda_c$ increases as we apply MF-$N({n_i}^{m_i})$
closures at higher levels $N$. Correspondingly, if we interpret $\Lambda$ as an
inverse-temperature-like control parameter, we find that the critical
temperature decreases as the closure level increases. Ultimately, as we take
$N \to \infty$, we find that all possible mean-field closures converge toward 
the infinite-order result [Eq.\ (\ref{equ:phihinf})], and we have 
$\Lambda_c \to \infty$. 

%

\subsection{Exponential closures at finite order}
\label{sec:altc}

Within the standard-MCT framework, the microscopic (non-schematic) mode-coupling 
equations are usually solved using a MF-$2(1^2)$ mean-field closure. 
The main reason for this is that treatment
of the full hierarchy is generally an intractable problem. Mean-field solutions
constitute an approximation to the exact solution of the hierarchy.
Here we discuss the possibility to use an alternative closure for finding
approximate solutions and discuss its advantages.

We have seen above that as we close the hierarchy of the
Mayer-Miyazaki-Reichman model [Eq.\ (\ref{equ:MCTdef})] 
at increasingly higher levels $N$, the type of closure used becomes
irrelevant. In the limit $N \to \infty$ and under a weak boundedness constraint
there is a unique solution [Eq.\ (\ref{equ:phihinf})] for the full hierarchy of
GMCT equations. This is the basis for our freedom to propose alternative
closures $\phi_N(t) = \ccal(\{ \phi_n(t) \},t)$.
Let us consider the shape of the exact solutions
at high levels $n$ in order to get a clue for a good choice of the closure.
From Eq.\ (\ref{equ:phint}) solutions at all levels $n$ are of the form $\phi_n(t) 
= \sum_{k \geq 0} r_k \rme^{s_k t}$. Numerical analysis suggests that for
sufficiently large $n$ and at any fixed $\Lambda$ we have $s_k \to -n-k$ for
$k \geq 0$ and $r_k \to 0$ for $k \geq 1$ but $r_0 \to 1$, therefore
$\phi_n(t) \approx \rme^{-n t}$. But if we had $\phi_{n+1}(t) = 0$,
then $\phi_n(t) = \rme^{-n t}$ would actually be exact according to 
Eq.\ (\ref{equ:MCTdef}). This suggests that a natural closure for our
hierarchy of GMCT equations consists in terminating it at some finite level
$N$ by setting $\phi_{N+1}(t) = 0$. We will refer to this type of truncation
as an \textit{exponential}
closure exp-$N$, since it implies that $\phi_{N}(t) \sim e^{-N t}$. 

A great advantage of exp-$N$ closures is that it is almost trivial to derive the
corresponding solutions $\phi_n(t)$. We recall Eqs.\ (\ref{equ:contfrac})-(\ref{equ:recphi}),
which apply in general and for any closure. For the exp-$N$ case, where $\phi_N(t) 
= \rme^{-N t}$ and hence $\phih_N(s) = 1/(s+N)$, Eq.\ (\ref{equ:mapfrac})
defines the ratio $\cphi_{N+1}(s)/\cphi_N(s) = 1/(s+N)$. Because the $\cphi_n(s)$
are only determined up to an overall factor we may set $\cphi_{N+1}(s) = 1$ and
$\cphi_N(s) = s+N$. Via iteration of the recursion formula (\ref{equ:recphi})
the whole set $\{ \cphi_n(s) \} = \{ \cphi_1(s),\ldots,\cphi_{N+1}(s) \}$
follows. Clearly the $\cphi_n(s)$ are polynomials of degree $N+1-n$ in $s$.
While this implies that $Q(s) = \cphi_1(s)$ is a polynomial of degree $N$
one shows that $R(s) = \frac{1}{s} [\cphi_1(s) - \cphi_2(s)]$ is in fact a
polynomial of degree $N-1$. Therefore, according to Eq.\ (\ref{equ:mapfrac}),
\begin{equation}
  \phih_1(s) = \frac{R(s)}{Q(s)} = \sum_{k=0}^{N-1} \frac{r_k}{s-s_k}.
  \label{equ:phihPQ}
\end{equation}
Here we have assumed that the $N$ roots $\{s_k\}$ of $Q(s_k) = 0$ are distinct
and performed a partial fractions decomposition of $R(s)/Q(s)$; the coefficients
$\{ r_k \}$ satisfy $r_k = \mathrm{Res}(\phih_1(s),s \to s_k)$. Inversion
of the Laplace transform in Eq.\ (\ref{equ:phihPQ}) produces $\phi_1(t) = \sum_{k=0}^{N-1} 
r_k \, \rme^{s_k t}$ which is analogous to Eq.\ (\ref{equ:phint}), but with the infinite sum
replaced by a finite one. We order the roots $s_{k+1} < s_k$ as before such that
the slow mode in $\phi_1(t)$ is $r_0 \, \rme^{s_0 t}$ with alpha-relaxation time
$\taua = -1/s_0$.

\begin{figure}
        \begin{center}
    \epsfig{file=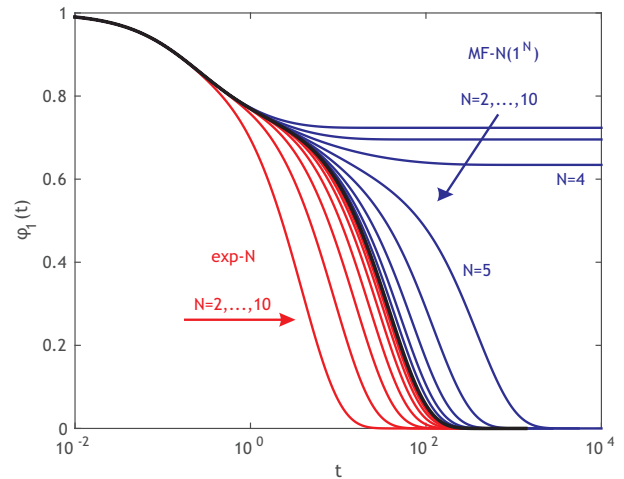,width=8cm,clip}
  \end{center}
  \caption{\label{fig:phi5} 
Solutions $\phi_1(t)$ for the Mayer-Miyazaki-Reichman model with $\Lambda=5$ 
under exp-$N$ (red curves) and MF-$N(1^N)$ (blue curves) closures. The black curve
  in the center is the solution of the infinite hierarchy.}
\end{figure}

Let us now investigate the results obtained from exp-$N$ closures. The simplest case
is exp-1 and gives $\phi_1(t) = \rme^{-t}$ independently of $\Lambda$. The first
non-trivial closure that predicts two-step relaxation is exp-2. Here $\phi_1(t) = 
r_0 \, \rme^{s_0 t} + r_1 \, \rme^{s_1 t}$ where $\{ s_0, s_1 \}$ are the roots of
$Q(s) = (s+1)(s+2) + \Lambda \, s$. At $\Lambda = 0$ we obviously have $s_0 = -1, s_1 = -2$
which gives $r_0=1, r_1 = 0$ and the exact solution $\phi_1(t) = \rme^{-t}$ is recovered.
In contrast to exp-1, however, the exp-2 solution starts to develop a plateau as
$\Lambda > 0$ is increased. The exp-2 alpha-relaxation time is given by
\begin{equation}
  \taua = \left[ \frac{3+\Lambda}{2} - \sqrt{\left( \frac{3+\Lambda}{2} \right)^2 -2} \right]^{-1}. 
  \label{equ:T2alpha}
\end{equation}
At small $\Lambda$ it grows like $\taua = 1 + \Lambda + \ocal(\Lambda^2)$, which
agrees to linear order in $\Lambda$ with the alpha-relaxation time for the infinite
hierarchy defined by Eq.\ (\ref{equ:sk}). However, at large $\Lambda$ the infinite hierarchy
yields essentially exponential growth of $\taua$ [Eq.\ (\ref{equ:arr})] while for the exp-2 closure
$\taua \sim \Lambda/2$ from Eq.\ (\ref{equ:T2alpha}). Numerical analysis of
Eq.\ (\ref{equ:phihPQ}) leads to a similar picture for exp-$N$ closures at higher levels
$N \geq 3$. With increasing closure level $N$ the range in $\Lambda$ over which
the exp-$N$ solutions give a good approximation to the exact result (\ref{equ:phint})
grows. At sufficiently large $\Lambda$, however, exp-$N$ closures always \textit{underestimate}
the alpha-relaxation time. Conversely, the solutions for MF-$N$
mean-field closures tend to \textit{overestimate} the alpha-relaxation time. We have illustrated
this in Fig.~\ref{fig:phi5}, which shows the solution $\phi_1(t)$ of the infinite
hierarchy at $\Lambda = 5$ and various MF-$N(1^N)$ and exp-$N$ solutions. Consider
first the mean-field solutions: for $N = 2,3,4$ we have $\Lambda_c < 5$ from
Fig.~\ref{fig:mfn} such that these MF-$N(1^N)$ solutions have $\taua = \infty$.
Only when we close the hierarchy at levels $N \geq 5$ does $\Lambda_c$ exceed 5
and the corresponding solutions $\phi_1(t)$ relax. According to the data in
Fig.~\ref{fig:phi5} the solution of the infinite hierarchy with $\Lambda = 5$
is well approximated by MF-$N(1^N)$ solutions with $N \geq 10$. Now, on the other
hand, consider the exp-$N$ solutions: these underestimate the alpha-relaxation
time and hence relax too quickly. From Fig.~\ref{fig:phi5} the exp-2 alpha-relaxation
time at $\Lambda = 5$ is by about a factor 10 too small. But as we raise the
closure level $N$ the exp-$N$ solutions quickly converge to the exact one.
According to Fig.~\ref{fig:phi5}, the exp-8 closure produces an approximation
just as good as the MF-$10(1^{10})$ one.

The data in Fig.~\ref{fig:phi5} summarizes various interesting features of our
hierarchy (\ref{equ:MCTdef}) of GMCT equations. Most importantly we have a unique
solution $\phi_1(t)$ for $N \to \infty$ regardless of the (bounded) closure.
Mean-field closures at finite order, such as the standard MF-$2(1^2)$ closure,
constitute one possibility to obtain approximations to the solution of the full
hierarchy. Simple truncation of the hierarchy represents another alternative
for obtaining such approximations. Regardless of the closure type, good approximations are only obtained
if the level of closure $N$ is chosen sufficiently large. With increasing coupling
$\Lambda$ between different levels of the hierarchy, more levels contribute to
$\phi_1(t)$, and hence the closure level $N$ must be increased for accurate results
over increasingly long time scales.
Whether one applies a mean-field closure or just terminates the hierarchy at
level $N$ only has a secondary effect on the result. From a pragmatic point of view,
truncation of the hierarchy appears to be the favorable option: on the one hand, it does not introduce extra
nonlinearities which makes an analytical approach possible, and on the other hand,
it also simplifies numerical methods because Eq.\ (\ref{equ:MCTdef}) is no longer
a self-consistency problem.


\section{More generic schematic models: continuous, discontinuous, and avoided transitions}
\label{sec:genericsGMCT}

The Mayer-Miyazaki-Reichman model discussed in the previous section represents
the simplest case of an infinite schematic GMCT hierarchy, with constant
coupling constants $\lambda_n=\Lambda$.  It is plausible, however, that
realistic glass-forming systems will conform to more intricate functional forms
of the $\lambda_n$ (and $\mu_n$) parameters, such as the models discussed in
Ref.\ \cite{janssen:14}. Here we turn our attention to these more generic
schematic GMCT hierarchies, which are characterized by e.g.\ a linear or
power-law growth of the $\lambda_n$ parameters as a function of level $n$. As
was already shown in Ref.\ \cite{janssen:14}, these models can give rise to a rich
pallet of glass transitions, ranging from sharp MCT-like (continuous or discontinuous)
to strictly avoided transitions, and with different types of alpha-relaxation-time and
plateau-height scaling behaviors.  In this section, we seek to provide more
insight into these models and identify general links between the
$\{\mu_n,\lambda_n\}$ parameters and the type of glass transition. Specifically,
we will focus on infinite hierarchies of the form (\ref{eq:sGMCThierarchy}) with
parameters
\begin{gather}
 \left\{ \mu_n=n, \lambda_n=\Lambda(n+c); c\geq0 \right\}, 
      \label{lin_model} \\
 \left\{ \mu_n=n, \lambda_n=\Lambda n^{1-\nu}; \nu>0 \right\}, 
      \label{pow_model} \\
 \left\{ \mu_n=\prod_{i=1}^{n-1}\frac{(2i+2)(i+b)}{(2i-1)(i+a)}, 
   \lambda_n=\Lambda \mu_n \frac{(2n+2)}{2n-1} \right\}. 
      \label{F2_model}
\end{gather}
The latter model with $a\approx 0.527\,26$ and $b\approx -0.237\,72$ has been
chosen such that it numerically reproduces the results of the $F_2$ model,
i.e., Eq.\ (\ref{F2_model}) represents a numerically accurate mapping of an
\textit{infinite} schematic GMCT hierarchy onto the 2-level hierarchy of Leutheusser
\cite{janssen:14}. We will refer to this infinite hierarchy as ''$F_2$-GMCT".
Also note that the Mayer-Miyazaki-Reichman model of Ref.\
\cite{mayer:06} is a special case of the hierarchy (\ref{pow_model}) with
$\nu=1$.

\subsection{General convergence of the hierarchy}
\label{sec:genconv}

We first address the convergence pattern of the models
(\ref{lin_model})-(\ref{F2_model}) as a function of closure level $N$. Unlike
the Mayer-Miyazaki-Reichman model, an analytic time-dependent solution for
these more generic hierarchies is not at hand, but it is fairly straightforward
to generate numerical results for $\phi_n(t)$ up to arbitrary order. By
considering both mean-field closures MF-$N$ and truncations exp-$N$, we can
investigate numerically how the predicted dynamics for $\phi_1(t)$ evolves over
time as $N$ increases.

\begin{figure}[h!]
        \begin{center}
    \epsfig{file=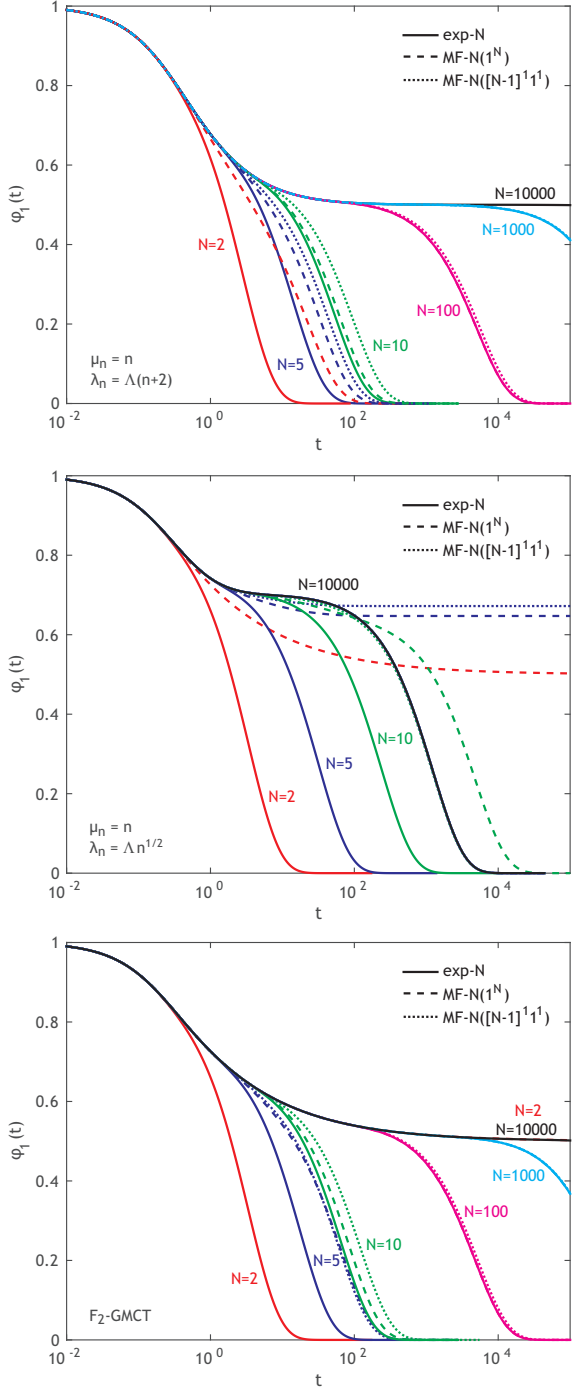,width=7.5cm}
  \end{center}
  \caption{
  \label{fig:conv_genmodels} The solutions $\phi_1(t)$ for different schematic
  GMCT models under both exponential and mean-field closures. The top panel
  shows the results for the model of Eq.\ (\ref{lin_model}) for $\Lambda=1$ and $c=2$,
  the middle panel shows the hierarchy of Eq.\ (\ref{pow_model}) for $\Lambda=4$
  and $\nu=1/2$ (here the data for $N=100$ and $N=1000$ are indistinguishable from $N=10000$),
  and the lower panel shows the results for the $F_2$-GMCT hierarchy
  of Eq.\ (\ref{F2_model}) for $\Lambda=1$. Note that the latter model is designed such
  that the infinite-order solution is identical to its MF-2($1^2$) closure.
  }
\end{figure}

In Fig.\ \ref{fig:conv_genmodels}, we show representative examples for the
generic hierarchies of Eqs.\ (\ref{lin_model})-(\ref{F2_model}) for closure
levels $N=2$, 5, 10, 100, 1000, and 10000. The latter serves as our numerical
representation of the infinite-order result. It may be seen that the
exponential closures always yield a \textit{lower} bound to the dynamics,
analogous to what was found for the Mayer-Miyazaki-Reichman model.  This can be
understood by considering that a truncation of the hierarchy rigorously
removes the memory effects from $n > N$, resulting in faster decay for the
$\phi_n(t)$ at the lower-lying levels. For any given mean-field closure level $N$,
we find that the MF-N$((N-1)^1 1^1)$ closures generally follow the
infinite-order solution more closely than the MF-N$(1^N)$ results. This is also
consistent with the observations made for the Mayer-Miyazaki-Reichman model.
However, Fig.\ \ref{fig:conv_genmodels} indicates that the mean-field closures
do not always give an upper bound to the dynamics. 
For example, for the hierarchy
of Eq.\ (\ref{lin_model}) with $c=2$, the MF-2($1^2$) closure gives relatively
fast decay, while the infinite-order solution predicts an ergodicity-breaking
transition with a $\phi_1(t)$ that remains finite for all times $t$. Increasing
the mean-field closure level does systematically yield slower relaxation, and
ultimately reproduces the infinite-order result as $N$ tends to infinity. For
the hierarchy of Eq.\ (\ref{pow_model}) with $\nu=1/2$, the MF-$N$ convergence
pattern is more complex: the lowest-order MF-2($1^2$) closure predicts a sharp
transition at $\Lambda_c=4$, with a plateau height of $\phi_1(t\rightarrow \infty) =
0.5$.  Note that this case is completely identical to the $F_2$ model. As $N$
increases, the plateau height first increases (shown here for MF-5), thus implying a more
strongly arrested glass state. For sufficiently large $N$, however, the MF-$N$
closure brings the system back into the ergodic phase (shown here for MF-10),
and yields faster relaxation as $N$ increases. Ultimately, the series of MF-$N$
closures converges onto the infinite-order solution, which in the present
example occurs for MF-100.  The $F_2$-GMCT hierarchy of Eq.\ (\ref{F2_model})
represents an even more extreme case: since this model is \textit{designed} to
reproduce the MF-2($1^2$) result (i.e., the $F_2$ model) in the infinite-order
limit, the MF-2($1^2$) closure is numerically exact, and any higher
finite-order closure will yield only an approximate solution. In fact, as is
evident from the lower panel of Fig.\ \ref{fig:conv_genmodels}, any MF-$N$
closure with $N>2$ yields a \textit{lower} bound to the exact solution of the
$F_2$-GMCT model, similar to the exponential closure series.
Overall, these results demonstrate that mean-field closures of
infinite GMCT hierarchies do not constitute a rigorous upper bound to
the exact solution; the Mayer-Miyazaki-Reichman model of the previous section, for
which MF-$N$ closures do always yield an upper bound, is only a special case.

There is, however, also an important general convergence pattern that emerges from the data
in Fig.\ \ref{fig:conv_genmodels}, namely: \textit{regardless of the type of closure, an
increasing closure level yields results that are accurate (i.e., agree with the full hierarchy)
over longer time scales.}
More explicitly, for the examples considered here,
the shortest time scales up to $\ocal(1)$ are already well reproduced by an $N=2$
closure, an $N=10$ closure gives accurate results up to times $\ocal(10)$, 
$N=100$ closures give quantitative accuracy up to times $\ocal(10^2)$,
and $N=1000$ closures follow the infinite-order solution up to $\ocal(10^4)$.
In other words, rather than setting an upper or lower bound to the true dynamics,
the closure level implicitly sets a maximum on the time range that can be accurately
described. 
This is a general trend that is observed across the various
models considered here, and implies that a perturbative approach is warranted:
by systematically incorporating more higher-order contributions, the predicted dynamics
systematically converges, over increasingly long time scales, upon a single
infinite-order solution that is exact up to infinite times.
We note that a similar notion of convergence was also recently found in microscopic GMCT
calculations of a quasi-hard sphere system, which could be performed up to
closure level $N=4$ with all relevant wavevectors included \cite{janssen:15}.
 Altogether, these results provide compelling evidence that GMCT hierarchies are
\textit{generally convergent}, even though no small parameter exists in the
theory.

\subsection{Relation between the $\{\mu_n,\lambda_n\}$ parameters and the type of glass transition}
\label{sec:transtypes}

As a final part of our discussion, let us elaborate on how the choice of
$\{\mu_n,\lambda_n\}$ parameters governs the nature of the predicted glass
transition within the infinite-order framework. As already discussed in Ref.\
\cite{janssen:14} on general grounds, the asymptotic behavior of the
parameters in the $n\rightarrow \infty$ limit determines whether the transition
is sharp or avoided: if there is a sharp transition at a finite critical point
$\Lambda=\Lambda_c$, we have $\lim_{n\rightarrow \infty}
\lambda_n/\mu_{n+1} > 1$, so that the series for $\taua$ diverges at
$\Lambda_c$. Conversely, if the transition is rigorously avoided so that the
correlation functions ultimately decay to zero for all $\Lambda$, we have
$\lim_{n\rightarrow \infty} \lambda_n/\mu_{n+1} < 1$.  Furthermore, for sharp
transitions at finite $\Lambda_c$, we may distinguish between type-A and type-B
transitions, which are characterized by continuous and discontinuous growth of
the nonergodicity parameter $q_n$ [Eq.\ (\ref{equ:qndef})] at the critical point,
respectively.  Continuous transitions obey $\lim_{n\rightarrow \infty}
\mu_n/\lambda_n > 1$ at $\Lambda_c$ so that the $1/q_n$ series diverges, 
while discontinuous transitions have $\lim_{n\rightarrow \infty} \mu_n/\lambda_n < 1$.
Below we seek to provide more physical insight into these mathematical arguments.

It is important to first recall from Ref.\ \cite{janssen:14} that the infinite
hierarchies of Eqs.\ (\ref{lin_model}) and (\ref{F2_model}) give rise to a
sharp type-B transition at $\Lambda_c=1$.  Conversely, the model of Eq.\
(\ref{pow_model}) always yields an avoided transition with $\Lambda_c
\rightarrow \infty$, and the value of $\nu$ ($\nu>0$) determines whether the
corresponding relaxation time grows in an Arrhenius, sub-Arrhenius, or
super-Arrhenius fashion as a function of the control parameter $\Lambda$.
Hence, $\nu$ may be interpreted as a fragility parameter \cite{janssen:14}.
Note that the case $\nu=0$ would recover the hierarchy (\ref{lin_model}) with
$c=0$, which thus represents a ''transitional" scenario between avoided and
sharp type-B transitions. 

\begin{figure}[h!]
        \begin{center}
\epsfig{file=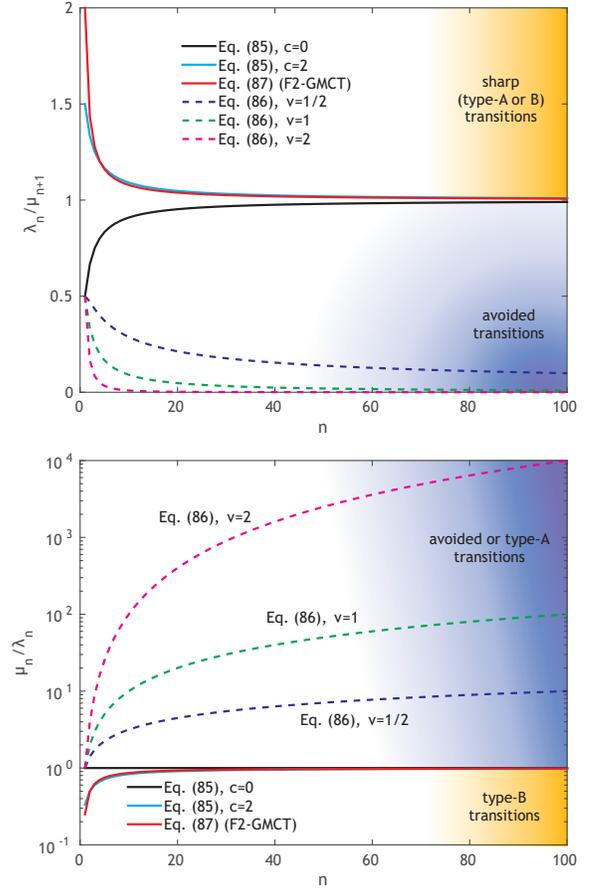,width=7.5cm}
  \end{center}
  \caption{
  \label{fig:params_genmodels}
   The ratios $\lambda_n/\mu_{n+1}$ (top) and $\mu_n/\lambda_n$ (bottom) for the
   various GMCT models (\ref{lin_model})-(\ref{F2_model}) as a function of level $n$.
   Solid lines correspond to models with a sharp transition, while dashed lines
   correspond to models with a strictly avoided transition. 
  }
\end{figure}

Figure \ref{fig:params_genmodels} shows the ratios $\lambda_n/\mu_{n+1}$ and $\mu_n/\lambda_n$ for the
various models (\ref{lin_model})-(\ref{F2_model}) as a function of $n$. For
simplicity we will restrict our discussion to the case $\Lambda=1$, which is
the critical point at which hierarchies (\ref{lin_model}) and (\ref{F2_model}) undergo a
discontinuous transition.  The plot of the ratio $\lambda_n/\mu_{n+1}$ 
immediately reveals the nature of sharp (type-A or type-B) glass
transitions: the coupling constants $\lambda_n$ remain of \textit{the same
order of magnitude} as the bare frequencies $\mu_n$ (or $\mu_{n+1}$) as $n$ tends towards infinity.  That is,
when a system is driven into the glass state through a sharp transition,
\textit{all} higher-order memory terms, up to $n\rightarrow \infty$, contribute
approximately equally at criticality (after normalizing with respect to the
bare frequencies). This, however, also raises a seeming paradox when considering
\textit{finite} hierarchies: the $F_2$ model, for example, which contains only an
$n=1$ contribution, with $\lambda_1=4\mu_1\Lambda$ and mean-field closure $\phi_2(t) = \phi_1(t)^2$,
also yields a sharp type-B transition at $\Lambda_c=1$. How can this be united with the
scenario of Fig.\ \ref{fig:params_genmodels} for the infinite $F_2$-GMCT hierarchy, which suggests that all $n>1$
play an important role in the $F_2$ model? The answer is simple: all $n>1$ contributions
in the $F_2$-GMCT model are necessary to ultimately yield a $\phi_2(t)$ \textit{that behaves 
numerically exactly as} $\phi_1(t)^2$. In other words, the $F_2$ model effectively 
captures all non-zero high-order ($1<n<\infty$) contributions of a non-trivial
infinite GMCT hierarchy into a single, simple MF-2($1^2$) closure ansatz. In this
sense, one may argue that the \textit{true} underlying physics of the $F_2$ model is 
in fact governed by infinitely many high-order correlations, rather than
by two-point density correlations alone.

From the foregoing discussion and the results in Fig.\ \ref{fig:params_genmodels}, it is also clear how
the $F_2$-GMCT model may be corrected to improve upon the predictions of the
$F_2$ model, which generally overestimates a system's tendency to form a glass.
By damping the higher-order $\lambda_n$ parameters, or increasing the $\mu_n$
at high $n$, the sharp transition can be converted into a weakly avoided
transition for which the asymptotic series $\lim_{n\rightarrow \infty}
\lambda_n/\mu_{n+1}$ becomes smaller than 1, and $\lim_{n\rightarrow \infty}
\mu_n/\lambda_n$ becomes larger than 1. Such corrections essentially make the
system more ''liquid-like", since they yield faster decay patterns for the
highest-$n$ correlation functions $\phi_n(t)$, which subsequently serve as
faster-decaying memory kernels for the lower-$n$ correlators.  Indeed, in the
case that we rigorously damp all high-order contributions by setting
$\lambda_n=0$ above a certain level $N$, we recover the exponential closure
exp-$N$ and ultimately get full temporal relaxation to zero
for all $\phi_n(t)$ at long times, thus removing the sharp transition.
Continuing this correction scheme up to the extreme case of an
exp-1 closure, for which \textit{all} higher-order
correlations rigorously vanish, we simply obtain fast exponential decay for
$\phi_1(t)$ and thus recover the normal-liquid regime. 

Finally, let us recall that the relative importance of the
higher-order correlations can also control the degree of fragility:
the smaller the value of $\nu$ in the hierarchy of Eq.\ (\ref{pow_model}),
the more fragile (or more
strongly $\Lambda$-dependent) the relaxation-time growth \cite{janssen:14}. 
Hence, it is plausible that glass-forming materials with different degrees
of fragility are governed by different effective strengths of the
higher-order memory terms. In the fully microscopic version of the theory,
this would have to be realized implicitly through the static structural input,
since the microscopic framework admits no free parameters. This would 
consequently provide a means to directly relate the structure to the dynamics.
The extent to
which the GMCT equations of Eqs.\ (\ref{eq:GMCTeqPhi_n})-(\ref{eq:Mn})
can account for different degrees of fragility on strictly first-principles
grounds, i.e., by taking only static structure factors as input, will
be tested in future work.

\section{Conclusions}
\label{sec:concl}

In summary, we have provided an extensive discussion of various models based on
the generalized mode-coupling theory of the glass transition. This
first-principles theory amounts to an infinite hierarchy of coupled equations
for dynamic multi-point density correlations, which can be made formally exact
if all relevant correlations are taken into account. The theory corrects upon
the well-established standard mode-coupling theory by delaying and ultimately
rigorously avoiding the \textit{ad hoc} MCT factorization of the dynamic
four-point correlator that governs the lowest-order memory kernel.  Within the
confines of \textit{schematic} infinite GMCT hierarchies, which ignore any
wavevector dependence, the theory greatly simplifies and becomes tractable both
analytically and numerically.  This simplification, however, necessitates the
introduction of new schematic parameters for the bare frequencies and the memory
kernels, which serve to capture, at least in a qualitatively sense, the full
wavevector-dependent expressions.

We have first discussed in detail the so-called Mayer-Miyazaki-Reichman model
of Ref.\ \cite{mayer:06}, which constitutes one of the simplest infinite
schematic GMCT models.  Remarkably, the infinite-order construct of this model
admits an analytic but highly non-trivial solution, which provides an excellent
benchmark for \textit{finite}-level closures of the hierarchy. It could be
shown rigorously that so-called mean-field closures, which factorize a
high-order multi-point correlation function in terms of lower-order ones (thus
essentially generalizing the standard-MCT factorization to arbitrary order),
uniformly converge upon the infinite-order solution as the closure level
increases. Such mean-field closures always approach the time-dependent
infinite-order solution from above, thus providing an upper bound to the exact
alpha-relaxation time.  Alternatively, one may also truncate the hierarchy at
an arbitrary level by simply setting the highest-order correlator to zero
(''exponential closure").  This approach leads to ultimately full relaxation of
all dynamic density correlation functions, and the corresponding
alpha-relaxation times always yield a lower bound to the exact result. As the
closure or truncation level tends to infinity, however, both closure types
systematically converge upon the exact infinite-order solution: \textit{the
more levels are included in the hierarchy, the longer the time scale that is
accurately reproduced.} Generally, one expects the mean-field closures to
converge faster onto the exact result in the strongly supercooled regime where
the dynamics is significantly slowed down, while the exponential closures
converge more rapidly in weakly supercooled and normal-liquid regimes.

Next, we have explored the more generic classes of infinite schematic GMCT
hierarchies of Ref.\ \cite{janssen:14}, which allow for more flexible functional
forms of the schematic parameters.
While an analytic infinite-order solution for these models is not available, we
have found numerically that all possible models considered here exhibit the
same uniform convergence pattern as the Mayer-Miyazaki-Reichman model:
regardless of the type of closure, the systematic inclusion of more
higher-order correlations yields results that quantatively reproduce
the dynamics of the infinite-order solution over systematically increasing times.
\textit{It thus appears that infinite GMCT hierarchies are
generally convergent, even though no small parameter exists in our theory that
would motivate such a perturbative approach.} This finding also provides hope for
fully microscopic GMCT calculations, for which it may be computationally
challenging to include many hierarchical levels. Indeed, we recently found \cite{janssen:15} that
full wavevector-dependent GMCT applied to a quasi-hard system up to closure
level $N=4$ yields a convergence pattern that is consistent with the
results of the present study.

Finally, we have sought to provide more insight into the role of higher-order
correlations by investigating how the various functional forms of the schematic
parameters affect the nature of the predicted dynamical glass transition. It
was found that sharp transitions, which occur at a finite value of the control
parameter, are governed by contributions from \textit{all} higher-order
contributions, up to the infinite-order level of the hierarchy. Conversely, for
avoided transitions the higher-order correlations become less important as the
hierarchical level increases.  A sharp transition may thus be rounded off and
converted into an avoided transition by reducing the relative contributions of
higher-order correlations.  Overall, these insights can help to gain a better
understanding of the physical  mechanisms that underlie the glass transition,
and allow for new means to elucidate the role of structure in dynamical arrest.

\acknowledgments
LMCJ thanks the Netherlands Organization for Scientific Research (NWO) for financial support
through a Rubicon fellowship. DRR ackowledges grant NSF-CHE 1464802 for support.


\begin{appendix}

\section{The functions $A_n(s), \ldots, D_n(s)$}
\label{sec:ABCD}

The functions $A_n(s), \ldots, D_n(s)$ were introduced in (\ref{equ:phiAD}). Here we summarize
relevant facts about them. First, their explicit form is
\begin{eqnarray}
  A_n(s) & = & \Phi(s+1,s+2;\Lambda) \Psi(s,s+n;\Lambda) \nonumber \\ 
  & & + \Gamma(n) \Phi(s,s+n;\Lambda) \Psi(s+1,s+2;\Lambda), 
  \label{equ:Ans} \\ 
  B_n(s) & = & s^{-1} [ \Phi(s,s+1;\Lambda) \Psi(s,s+n;\Lambda) \nonumber \\ 
  & & - \Gamma(n) \Phi(s,s+n;\Lambda) \Psi(s,s+1;\Lambda)], 
  \label{equ:Bns} \\ 
  C_n(s) & = & \Phi(s+1,s+2;\Lambda) \Psi(s+1,s+n+1;\Lambda) \nonumber \\ 
  & & - \Gamma(n) \Phi(s+1,s+n+1;\Lambda) \Psi(s+1,s+2;\Lambda), 
\nonumber \\ 
  \label{equ:Cns} \\ 
  D_n(s) & = & \Phi(s,s+1;\Lambda) \Psi(s+1,s+n+1;\Lambda) \nonumber \\ 
  & & + \Gamma(n) \Phi(s+1,s+n+1;\Lambda) \Psi(s,s+1;\Lambda). 
  \label{equ:Dns} 
\end{eqnarray}
The hypergeometric functions $\Phi$ and $\Psi$ appearing in these expressions are analytic with
respect to $s \in \mathbb{C}$. It therefore follows immediately that $A_n(s)$, $C_n(s)$ and $D_n(s)$
are analytic, too. Regarding $B_n(s)$ we will see below that the term in the square brackets
vanishes at $s = 0$. It is thus $\ocal(s)$ around $s=0$ from analyticity and a Taylor expansion.
Consequently the $s^{-1}$ singularity in (\ref{equ:Bns}) is lifted and $B_n(s)$ is likewise an
entire function of $s \in \mathbb{C}$.
The $s \to 0$ limits of $A_n(s), \ldots, D_n(s)$, which we denoted $A_n, \ldots, D_n$, were used
in Section~\ref{sec:mfn}. Expressions for $A_n$, $C_n$ and $D_n$ are readily obtained from
(\ref{equ:Ans}), (\ref{equ:Cns}) and (\ref{equ:Dns}). Based on the series expansions
(\ref{equ:Phiser}) and (\ref{equ:Psiser}) one verifies that $\Phi(0,n+1;\Lambda) = 1/n!$ and
$\Psi(0,n+1;\Lambda) = 1$ for integer $n \geq 0$. Similarly one shows that
\begin{eqnarray}
  \Phi(1,n+2;\Lambda) & = & \frac{1}{\Lambda^{n+1}} \left( \rme^\Lambda - \sum_{k=0}^n \frac{\Lambda^k}{k!} \right), 
  \label{equ:Phi1n} \\ 
  \Psi(1,n+2;\Lambda) & = & \frac{n!}{\Lambda^{n+1}} \sum_{k=0}^n \frac{\Lambda^k}{k!}. 
  \label{equ:Psi1n} 
\end{eqnarray}
By setting $s = 0$ in (\ref{equ:Ans}), (\ref{equ:Cns}) and (\ref{equ:Dns}) and using these identities
one immediately obtains $A_n$, $C_n$ and $D_n$ as given in (\ref{equ:An}), (\ref{equ:Cn}) and (\ref{equ:Dn}).
Equation (\ref{equ:Bns}), however, is less trivial to deal with since it is indeterminate, i.e.\ $0/0$, at
$s = 0$. We also want to avoid $s$-derivatives for taking the limit $s \to 0$ because these are no
hypergeometric functions anymore. Instead we integrate the identity~\cite{Mathbook}
$\partial_x \Phi(a,b;x) = a \, \Phi(a+1,b+1;x)$ 
to rewrite
\begin{equation}
  \Phi(s,s+n;\Lambda) = \frac{1}{\Gamma(s+n)} + s \int_0^\Lambda \rmd x \, \Phi(s+1,s+n+1;x). 
  \label{equ:PhiIntId}
\end{equation}
Here $\Phi(s,s+n;0) = 1/\Gamma(s+n)$ was used. When substituting this representation in (\ref{equ:Bns})
we find $B_n = B_n(s \to 0)$,
\begin{eqnarray}
  B_n & = & \lim_{s \to 0} \frac{1}{s} \bigg[ \frac{1}{\Gamma(s+1)} \Psi(s,s+n;\Lambda) \nonumber \\ 
&&    - \frac{\Gamma(n)}{\Gamma(s+n)} \Psi(s,s+1;\Lambda) \bigg] 
    \nonumber \\
  & & + \Psi(0,n;\Lambda) \int_0^\Lambda \rmd x \, \Phi(1,2;x)  \nonumber \\
&&    - \Gamma(n) \, \Psi(0,1;\Lambda) \int_0^\Lambda \rmd x \, \Phi(1,n+1;x). 
    \nonumber 
\end{eqnarray}
%
%
In the last two lines of this equation the $1/s$ factors canceled against the explicit $s$ from
(\ref{equ:PhiIntId}) such that we were able to take the limit $s \to 0$ immediately. The first two lines,
however, are still of the form $0/0$. Here we pull out an overall factor $\Gamma(n)/\Gamma(s+n)$ from
the square bracket -- which reduces to unity for $s \to 0$ -- and use $\Gamma(x+1) = x \, \Gamma(x)$
to simplify $\Gamma(s+n)/\Gamma(s+1)$,
\begin{eqnarray}
  B_n & = & \lim_{s \to 0} \frac{1}{s} \Bigg\{ \frac{1}{(n-1)!} 
    \Bigg[ \prod_{k=1}^{n-1} (s+k) \Bigg] \Psi(s,s+n;\Lambda) 
\nonumber \\
&& - \Psi(s,s+1;\Lambda) \Bigg\} 
    \nonumber \\ 
  & & + \Bigg[\int_0^\Lambda \rmd x \, \Phi(1,2;x) - (n-1)! 
\nonumber \\
&& \times \int_0^\Lambda \rmd x \, \Phi(1,n+1;x) \Bigg]. 
\end{eqnarray}
%
%
The factors $\Psi(0,1;\Lambda)$, $\Psi(0,n;\Lambda)$ in the last two lines dropped out since they equal
unity as discussed above. For taking $s \to 0$ we now express $\Psi(s,s+n;\Lambda)$ via (\ref{equ:Psiser}),
where in particular $\Psi(s,s+1;\Lambda) = \Lambda^{-s}$, and remove the overall factor $\Lambda^{-s} \to 1$
for $s \to 0$. 
The curly brackets then just contain a polynomial in $s$ with a root
at $s = 0$ and hence the $s \to 0$ limit is straightforward. We further simplify $B_n$ using
\begin{eqnarray*}
  \int_0^\Lambda \rmd x \, \Phi(1,n+1;x) &=& 
  - \frac{\Lambda}{n-1} \, \Phi(1,n+1;\Lambda)  \nonumber  \\
&& + \frac{1}{n-1} \int_0^\Lambda \rmd x \, \Phi(1,n;x). 
\end{eqnarray*}
%
%
One verifies this identity by substituting (\ref{equ:Phi1n}) and an integration by parts. Applying it
recursively in the latter expression for $B_n$ eventually leads to cancellation of the non-elementary
integral $\int_0^\Lambda \rmd x \, \Phi(1,2;x)$. Altogether one arrives at the expression (\ref{equ:Bn})
for $B_n$ given in the main text.

\end{appendix}


\end{document}